# Imaging spinon density modulations in a 2D quantum spin liquid

Wei Ruan[1,2,†], Yi Chen[1,2,†], Shujie Tang[3,4,5,6,7], Jinwoong Hwang[5,8], Hsin-Zon Tsai[1,10], Ryan Lee[1], Meng Wu[1,2], Hyejin Ryu[5,9], Salman Kahn[1], Franklin Liou[1], Caihong Jia[1,2,11], Andrew Aikawa[1], Choongyu Hwang[8], Feng Wang[1,2,12], Yongseong Choi[13], Steven G. Louie[1,2], Patrick A. Lee[14], Zhi-Xun Shen[3,4], Sung-Kwan Mo[5], Michael F. Crommie[1,2,12,*]

*[1]Department of Physics, University of California, Berkeley, California 94720, USA*

*[2]Materials Sciences Division, Lawrence Berkeley National Laboratory, Berkeley, California 94720, USA*

*[3]Stanford Institute for Materials and Energy Sciences, SLAC National Accelerator Laboratory and Stanford University, Menlo Park, California 94025, USA*

*[4]Geballe Laboratory for Advanced Materials, Departments of Physics and Applied Physics, Stanford University, Stanford, California 94305, USA*

*[5]Advanced Light Source, Lawrence Berkeley National Laboratory, Berkeley, California 94720, USA*

*[6]CAS Center for Excellence in Superconducting Electronics, Shanghai Institute of Microsystem and Information Technology, Chinese Academy of Sciences, Shanghai 200050, China*

*[7]School of Physical Science and Technology, Shanghai Tech University, Shanghai 200031, China*

*[8]Department of Physics, Pusan National University, Busan 46241, Korea*

*[9]Center for Spintronics, Korea Institute of Science and Technology, Seoul 02792, Korea*

*[10]International Collaborative Laboratory of 2D Materials for Optoelectronic Science & Technology of Ministry of Education, Engineering Technology Research Center for 2D Material Information Function Devices and Systems of Guangdong Province, Shenzhen University, Shenzhen 518060, China*

*[11]Henan Key Laboratory of Photovoltaic Materials and Laboratory of Low-dimensional Materials Science, School of Physics and Electronics, Henan University, Kaifeng 475004, China*

*[12]Kavli Energy Nano Sciences Institute at the University of California Berkeley and the*

*Lawrence Berkeley National Laboratory, Berkeley, California 94720, USA*

[13]*Advanced Photon Source, Argonne National Laboratory, Argonne, Illinois 60439, USA*

[14]*Department of Physics, Massachusetts Institute of Technology, Cambridge MA 02139, USA*

*† These authors contributed equally to this work.*

**e-mail: crommie@berkeley.edu*

**Abstract**

Two-dimensional triangular-lattice antiferromagnets are predicted under some conditions to exhibit a quantum spin liquid ground state whose low-energy behavior is described by a spinon Fermi surface. Directly imaging the resulting spinons, however, is difficult due to their fractional, chargeless nature. Here we use scanning tunneling spectroscopy to image spinon density modulations arising from a spinon Fermi surface instability in single-layer 1T-$TaSe_2$, a two-dimensional Mott insulator. We first demonstrate the existence of localized spins arranged on a triangular lattice in single-layer 1T-$TaSe_2$ by contacting it to a metallic 1H-$TaSe_2$ layer and measuring the Kondo effect. Subsequent spectroscopic imaging of isolated, single-layer 1T-$TaSe_2$ reveals long-wavelength modulations at Hubbard band energies that reflect spinon density modulations. This allows direct experimental measurement of the spinon Fermi wavevector, in good agreement with theoretical predictions for a 2D quantum spin liquid. These results establish single-layer 1T-$TaSe_2$ as a new platform for studying novel two-dimensional quantum-spin-liquid phenomena.

Since the first predictions by Anderson in 1973[1], quantum spin liquids (QSLs) have been intensely investigated[2-7] as candidates to host exotic quantum phenomena such as fractionalized elementary excitations[2-4], topological order[4,5], and unconventional superconductivity[6,7]. QSLs are a novel state of matter predicted to arise in quantum antiferromagnets where geometric frustration and quantum fluctuations are strong enough to prevent a magnetically ordered ground state[2-4]. A key to understanding the QSL state is its low-energy physics, often dominated by emergent fractional fermions (termed spinons) which carry spin-1/2 but no charge[2-4]. Many QSL models are based on two-dimensional (2D) triangular-lattice Mott insulators[2-4], and several material candidates incorporating coupled 2D layers have been found to exhibit behavior consistent with spinon excitations[8-10]. It remains debatable whether the spinons in such systems are gapped or not[4,8-10], but increasing evidence[10] suggests the existence of gapless spinons that exhibit a Fermi surface (FS)[2-4,8-13].

Imaging the itinerant spinons, however, has been challenging due to their fractional and chargeless nature. Some predictions suggest that the spinon FS should host instabilities leading to spinon spatial patterns[13] that can, in principle, be imaged by single-particle scanned probes[14]. Bulk 1T-$TaS_2$ has been suggested as one such QSL candidate[15-21]. This layered material is believed to exhibit a Mott insulator ground state that arises cooperatively from a star-of-David charge density wave (CDW)[22-24]. Each star-of-David cluster contributes one localized spin, thus forming a triangular spin lattice. Some evidence for QSL behavior in bulk 1T-$TaS_2$ has been found, such as the absence of magnetic order down to millikelvin temperature[17-19] and a linear term in the thermal conductivity[20,21], but QSL physics here is complicated by interlayer coupling. This is because interlayer coupling can lead to spin delocalization and/or interlayer spin-singlet formation, which are detrimental to forming a gapless QSL state[15,19,25-27].

Here we report the observation of a QSL exhibiting a spinon FS in single-layer (SL)

$1T\text{-}TaSe_2$ through the use of scanning tunneling microscopy/spectroscopy (STM/STS). SL $1T\text{-}TaSe_2$ is a newly discovered 2D Mott insulator that exhibits a low-temperature star-of-David CDW phase similar to bulk $1T\text{-}TaS_2$ (i.e., the star-of-David cells are centered at Ta atoms - Supplementary Fig. 1), but which does not suffer from the disadvantages of interlayer coupling between 1T layers[28,29]. We report two new findings that support the presence of a QSL in SL $1T\text{-}TaSe_2$. First we demonstrate the existence of localized spins on a triangular lattice in SL $1T\text{-}TaSe_2$ through the observation of a Kondo resonance at the Fermi level ($E_F$) when SL $1T\text{-}TaSe_2$ is placed in contact with metallic $1H\text{-}TaSe_2$. Next we show evidence for a QSL-based spinon FS subject to a FS instability in isolated SL $1T\text{-}TaSe_2$ through the observation of long-wavelength modulations in the electronic local density of states (LDOS) at the Hubbard band energies. Such wave patterns are unexpected in an insulator but naturally occur in the presence of a spinon FS. The ability to access fractional spinon behavior via a single-particle probe (i.e., STM) arises from the decomposition of injected electrons into spinons (chargeless spin-1/2 fractional particles) and chargons (spinless charged fractional particles)[14,30]. By imaging spinon-induced LDOS modulations we are able to directly determine the spinon Fermi wavevector ($k_F$), in good agreement with theoretical predictions[15,16].

**Kondo resonance in a 1T/1H $TaSe_2$ vertical heterostructure**

SL $TaSe_2$ films were grown on epitaxial bilayer graphene (BLG) terminated 6H-SiC(0001) substrates and also on cleaved graphite surfaces via molecular beam epitaxy (MBE)[29,31]. A single layer of $TaSe_2$ contains one Ta atomic layer sandwiched between a pair of Se atomic layers, with each Ta atom coordinated by six Se atoms (Fig. 1a). The Se cage forms an octahedron in the metastable 1T-phase and a trigonal prism in the stable 1H-phase. Coexisting 1T and 1H phases were grown via MBE in the SL limit under controlled growth conditions, as shown by our STM images (Fig. 1b). Atomically-flat 1T and 1H SL islands can easily be distinguished as seen by close-up images of both phases in the insets to Fig. 1b which exhibit a triangular $\sqrt{13} \times \sqrt{13}$ star-of-David CDW pattern for the 1T-phase[29] and a 3 × 3 CDW pattern for the 1H-phase[31]. Vertical heterostructures composed of a single 1T layer on top of a single 1H layer (1T/1H) are observed to display the star-of-David CDW pattern in the top layer and the 3 × 3 CDW in the bottom layer (Fig. 1b and Supplementary Fig. 2b).

We verified the electronic structure of the different $TaSe_2$ phases by measuring STM differential conductance (d$I$/d$V$) (which reflects surface electronic LDOS) as a function of sample bias voltage ($V_b$). d$I$/d$V$ spectra acquired on metallic SL 1H-$TaSe_2$ islands show finite LDOS at $E_F$ accompanied by a slight suppression due to CDW formation (Fig. 2a green curves), consistent with previous measurements[31]. d$I$/d$V$ spectra acquired on insulating SL 1T-$TaSe_2$ islands on BLG/SiC, on the other hand, show a Mott gap (Fig. 2a red curves), also consistent with previous measurements[29]. Here the LDOS peak near $V_b$ = -0.3 V is identified as the lower Hubbard band (LHB) and the upper Hubbard band (UHB) corresponds to the peaks near $V_b$ = 0.2 V (UHB$_1$) and $V_b$ = 0.6 V (UHB$_2$) (these are split due to reduced screening in 2D[29]). The electronic structure of the SL 1T and 1H phases was also identified using angle-resolved photoemission spectroscopy (ARPES) (Supplementary Fig. 3b), confirming the coexistence of both phases.

d$I$/d$V$ spectra acquired on 1T/1H vertical heterostructures are very different from either of the bare SL spectra at the Hubbard band energy scale. The heterostructure spectra reveal a pronounced zero bias peak (ZBP) that is absent from single layers (Fig. 2a blue curves) and which cannot be explained by doping[32-35] or strain (Supplementary Note 2.1). The ZBP was found to persist in every star-of-David cell (Supplementary Fig. 2). We identify the ZBP as a Kondo resonance[36,37], which is expected to arise when a local spin (from the 1T layer) couples to itinerant electrons in a metal (from the 1H layer)[38]. To test this hypothesis we examined the ZBP feature over the temperature range 5 K $\leqslant T \leqslant$ 70 K, as plotted in Fig.

2b. The ZBP gradually broadens with increased temperature by an amount that cannot be accounted for by pure thermal broadening (Supplementary Fig. 4a), but instead is well-fit by a thermally-convolved Fano line shape (Supplementary Fig. 4b) with an intrinsic temperature-dependent width $\Gamma(T)$. The intrinsic resonance width as a function of temperature ($T$) is observed to follow the well-known Kondo expression[37]

$$\Gamma(T) = \sqrt{(\pi k_{\mathrm{B}}T)^2 + 2(k_{\mathrm{B}}T_{\mathrm{K}})^2} \qquad (1)$$

where $k_{\mathrm{B}}$ is the Boltzmann constant, yielding an estimated Kondo temperature $T_{\mathrm{K}} = 57 \pm 3$ K (Fig. 2c, red dashed line) and a Kondo-coupling $J_{\mathrm{K}} \approx 0.2$ eV (Supplementary Note 2.1). The observation of the Kondo resonance peak in every star-of-David CDW cell suggests the existence of localized spins in isolated SL 1T-$TaSe_2$ that are arranged in a triangular lattice (Fig. 1c) (the coherence temperature of the 1T/1H Kondo lattice is lower than the STM base temperature of 5 K (Supplementary Note 2.2)). Measuring magnetism in single-layer materials via more direct methods is challenging. For example, we attempted to probe magnetism in SL 1T-$TaSe_2$ using X-ray magnetic circular dichroism but this yielded no observable magnetization beyond the noise level (Supplementary Fig. 5).

**Super-modulations in single-layer 1T-$TaSe_2$**

We explored QSL behavior in isolated single layers of 1T-$TaSe_2$ (i.e., supported by BLG/SiC rather than the metallic 1H phase) by characterizing long-wavelength modulations in the real-space electronic structure of this triangular spin lattice. Measurements were performed on SL 1T-$TaSe_2$ islands like the one shown in Fig. 3a which exhibit d$I$/d$V$ spectra characteristic of a Mott insulator (Fig. 3b). Fig. 3c-g shows constant-height d$I$/d$V$ maps acquired at different energies in the area outlined by a yellow dashed square in Fig. 3a, while Fig. 3h-l reveals the corresponding Fourier transform (FT) of each d$I$/d$V$ map (see Supplementary Fig. 6 for FT images without labels). The star-of-David CDW pattern dominates the empty-state LDOS for energies above the UHB[29] as seen by the d$I$/d$V$ map at $V_b$ = 1.0 V in Fig. 3c and its FT in Fig. 3h. Here the CDW triangular lattice yields 6-fold symmetric FT peaks marked by red circles and labeled as the primary reciprocal lattice vectors $\mathbf{b}_i$ ($1 \leqslant i \leqslant 6$). These define the first Brillouin zone (BZ) of the star-of-David CDW (red hexagon). No other periodicities are seen at this energy.

New longer-wavelength modulations emerge at lower energies closer to the Hubbard band energies (i.e., $UHB_2$, $UHB_1$, and LHB). At energies near $UHB_2$ ($V_b$ = 0.62 V), for example, a new super-modulation is clearly seen in the real-space image of Fig. 3d as bright patches of enhanced LDOS (see also Supplementary Fig. 7d). This structure corresponds to a triangular grid rotated by 30° from the CDW lattice with an incommensurate lattice constant slightly larger than $\sqrt{3}a$ (Supplementary Fig. 7) where $a$ is the CDW lattice constant. This incommensurate super-modulation (ICS) is best seen in Fig. 3i (the FT of Fig. 3d) which shows new peaks marked by blue circles along the Γ-K directions inside the CDW BZ. The ICS wavevector, $\mathbf{q}_{\text{ICS}}$, can be written in terms of the CDW reciprocal lattice vectors as $\mathbf{q}_{\text{ICS}} = q_{\text{ICS}}(\mathbf{b}_1 + \mathbf{b}_2)$ where $q_{\text{ICS}} = 0.241 \pm 0.009$ ($q_{\text{ICS}}$ is obtained from Gaussian fits of FTs acquired on 10 different islands with 5 different tips - Supplementary Fig. 8).

The ICS pattern is also observed when energy is lowered further to $UHB_1$, as seen by the d$I$/d$V$ map at $V_b$ = 0.2 V (Fig. 3e) and corresponding FT (Fig. 3j). The ICS is often obscured in real space images due to mixing with the CDW pattern (Supplementary Note 1), but its FT peaks are easily resolved from the CDW wavevectors (Fig. 3j). The ICS persists as energy is lowered to the LHB as seen in the d$I$/d$V$ map at $V_b$ = -0.18 V (Fig. 3f) which yields FT peaks at $\mathbf{q}_{\text{ICS}}$ (Fig. 3k). The filled-state LHB measurement, however, differs from the empty-state UHB measurements in that it also exhibits a short-range commensurate super-modulation (CS) of wavelength $\sqrt{3}a$ that yields broad FT peaks at the K-points of the CDW BZ (green circles in Fig. 3k - see also Supplementary Fig. 9). Both the ICS and CS disappear at energies below the LHB where only the star-of-David CDW remains as shown by the

d$I$/d$V$ map at $V_b$ = -0.8 V (Fig. 3g) and its corresponding FT (Fig. 3i).

The complete energy dependence of the super-modulations over the energy range -1V < $V_b$ < 1.5 V is shown in Fig. 4a and b (Supplementary Figs. 10 and 11 show additional data at selected energies). The energy-dependent FT amplitude along the Γ-K direction (black dashed line in the inset to Fig. 4a) as a function of wavevector $\mathbf{q} = q(\mathbf{b}_1 + \mathbf{b}_2)$ shows three main features: (i) the CDW reciprocal lattice vector $\mathbf{b}_1 + \mathbf{b}_2$ (at $q$ = 1) over a wide energy range (black dashed ovals), (ii) the CS (at $q$ = 1/3) in the filled-state LHB (green dashed oval), and (iii) the ICS (at $q = q_{\mathrm{ICS}} \approx 1/4$) in both the LHB and UHB regimes (blue dashed ovals). We observe that the ICS wavevector is independent of energy, but its *amplitude* is not. To better visualize the energy dependence of the ICS amplitude, its strength is defined as the ICS FT peak amplitude $A_{\mathrm{ICS}}$ normalized by the CDW FT peak amplitude $A_{\mathrm{CDW}}$. Fig. 4b shows that this ratio, $A_{\mathrm{ICS}}/A_{\mathrm{CDW}}$ (blue dots), is vanishingly small at all energies except for the LHB and UHB regimes (a SL 1T-$TaSe_2$ d$I$/d$V$ spectrum is plotted for reference (black line)).

In order to investigate possible substrate effects on the behavior of the observed super-modulation, SL 1T-$TaSe_2$ was also grown on cleaved graphite (HOPG) via MBE and characterized by STM/STS. These samples also exhibit star-of-David CDW order and Mott insulating behavior similar to the samples grown on BLG/SiC as shown by the STM images and d$I$/d$V$ spectra of Supplementary Fig. 12. STM images acquired in the LHB of SL 1T-$TaSe_2$/HOPG and their FTs show the same ICS pattern as seen for SL 1T-$TaSe_2$/BLG/SiC (Supplementary Fig. 12d and e). However, an additional 2 × 2 super-modulation (with respect to the CDW lattice) is sometimes seen for SL 1T-$TaSe_2$/HOPG in the LHB that has FT peaks near the M points (Fig. 5d and Supplementary Fig. 12f-k) (this was seen for 2 out of 11 islands). We refer to this new super-modulation wavevector as $\mathbf{q}_{\mathrm{M}}$. $\mathbf{q}_{\mathrm{ICS}}$ and $\mathbf{q}_{\mathrm{M}}$ were never simultaneously observed on the same SL 1T-$TaSe_2$/HOPG island.

**Relationship between super modulations and QSL behavior**

Our experimental results support the hypothesis that SL 1T-$TaSe_2$ is a 2D QSL. The first piece of evidence is that SL 1T-$TaSe_2$ contains a triangular lattice of localized spins, an essential ingredient for a QSL. This evidence is provided by our observation of the Kondo effect, which implies that each star-of-David in SL 1T-$TaSe_2$ contains a single quantum spin. The second piece of evidence is the long-wavelength modulations that we observe in SL 1T-$TaSe_2$ via STS imaging. These periodicities lie at the precise energies and wavevectors expected for a QSL, as discussed below.

Before describing how these periodicities are explained by spinon density modulations in SL 1T-$TaSe_2$, however, we first rule out alternative explanations for the observed super-modulations. The first alternative possibility is electronic quasiparticle interference (QPI). Our observed ICS pattern is inconsistent with QPI because its wavelength is energy independent over the Hubbard bandwidths. There are also no clear electronic structure features at the Hubbard band edges that would cause the QPI to be dominated by contributions at $\mathbf{q}_{ICS}$ (Supplementary Fig. 13). The second alternative possibility is a structural distortion such as a Moiré pattern and/or a surface reconstruction. This explanation is unlikely because the observed super-modulations exist only over very specific energy ranges involving the Hubbard bands, suggesting that they are electronic and/or spin-based phenomena. To further exclude Moiré patterns involving complex BLG/SiC reconstructions we point to the fact that SL 1T-$TaSe_2$/HOPG (HOPG has no reconstructions) exhibits the same ICS pattern as SL 1T-$TaSe_2$/BLG/SiC (Supplementary Fig. 12). We additionally see no evidence of the super-modulation in low-energy electron diffraction (LEED) patterns[29], and no evidence that it is a strain effect (Supplementary Note 3.1). The last alternative possibility we will mention is Peierls-type or correlation-driven charge order. The former typically requires an electron Fermi surface while the latter usually appears at much lower energy scales than the Hubbard bands[32]. Such behavior is inconsistent with the ICS observed in our system and has no reasonable connection to the new periodicities seen here (i.e., $\mathbf{q}_{\mathrm{ICS}}$ and $\mathbf{q}_{\mathrm{M}}$).

The presence of spinon density modulations, on the other hand, very naturally explains both the energy-dependence and wavevectors of the $\mathbf{q}_{\mathrm{ICS}}$ and $\mathbf{q}_{\mathrm{M}}$ super-modulations that we observe in SL 1T-$TaSe_2$ (Supplementary Note 3). The starting point here is the existence of a spinon FS for a triangular spin lattice that is subject to a spinon FS instability. The existence of a spinon FS within the CDW BZ can be intuitively understood from a simple counting argument since each star-of-David contains one unpaired spin and the total number of spinons (which are fermions) is equal to the total number of spins (thus leading to a half-filled spinon band). This is also supported by density matrix renormalization group (DMRG) calculations based on the triangular-lattice *t-J* model with a ring exchange term to simulate a Hubbard model near the Mott transition[16]. These calculations show that the spin-correlation peaks in the Γ-M directions are missing[16], thus suggesting a partial gap opening at the spinon Fermi wavevector $k_F$ along the Γ-M directions due to FS instabilities (Fig. 5a). The spinon $k_F$ in this direction can be theoretically estimated by modeling the spinon band structure with a zero-flux mean-field tight-binding Hamiltonian (Supplementary Fig. 14), yielding the value $k_F \approx 0.375$ reciprocal lattice units (r.l.u.).

The spinon FS instability wavevector $2k_F$ can be folded back into the first CDW BZ (Supplementary Fig. 15a) to yield $\mathbf{P}_i = (1 - 2k_{\mathrm{F}})\mathbf{b}_i \approx 0.249\mathbf{b}_i$ as marked by red dots in Fig. 5b. These primary $\mathbf{P}_i$'s do not coincide with our observed modulation wavevectors, but their higher harmonics (Fig. 5b and Supplementary Fig. 15) match well with the experimental super-modulations. For example, the theoretical harmonics $\mathbf{Q}_i = \mathbf{P}_i + \mathbf{P}_{i+1} \approx 0.249(\mathbf{b}_i + \mathbf{b}_{i+1})$ match our experimentally measured $\mathbf{q}_{\mathrm{ICS}} = (0.241 \pm 0.009)(\mathbf{b}_i + \mathbf{b}_{i+1})$ (Fig. 5c) while the theoretical harmonics $\mathbf{Q}_i' = 2\mathbf{P}_i$ match our experimentally measured $\mathbf{q}_{\mathrm{M}}$ (Fig. 5d). The experimental observation of predicted harmonics at both $\mathbf{Q}_i$ and $\mathbf{Q}_i'$ strongly suggests the existence of a hidden primary wavevector $\mathbf{P}_i$ and points to a spinon FS subject to instability at $\mathbf{P}_i$ in SL 1T-$TaSe_2$. The measured value of $\mathbf{q}_{\mathrm{ICS}}$, corresponding to the second harmonic of the folded $2k_F$ spinon Fermi wavevectors, thus yields an experimental value of $k_{\mathrm{F}} = (1 - q_{\mathrm{ICS}})/2 = (0.380 \pm 0.005)$ r.l.u., consistent with the theoretically predicted value of 0.375 r.l.u. This value of $k_{\mathrm{F}}$ is also consistent with our ARPES data which shows enhanced intensity centered at Γ that greatly weakens for in-plane momenta beyond $k_F$ (Supplementary Fig. 3).

The theoretical explanation of spinon density modulations at the periodicities observed experimentally is thus established, but a natural question is how an STM, which injects charged particles (i.e., electrons) into a material, can observe particles that have no charge (i.e., spinons). The answer lies in the process of fractionalization, whereby strongly correlated electrons in a QSL are predicted to separate into spinons (which have spin and no charge) and chargons (which have charge and no spin). Due to constraints in the total occupation of spinons and chargons, modulation of the spinon density can induce a small density modulation of physical charge[14]. However, an even stronger effect occurs in the tunneling process since an electron injected into the QSL UHB (or removed from the LHB) will fractionalize into (or recombine from) a spinon and a chargon (see sketch in Fig. 4c and d). The tunneling probability thus depends on both the spinon wavefunction and the chargon wavefunction (similar concepts have been used previously to interpret ARPES data[30,39]). Spatially periodic spinon density due to a spinon FS[11-13,16] should thus modulate quasiparticle tunnel rates at the Hubbard band energies (Supplementary Note 3.4 and Supplementary Fig. 16). Our experimental observation of super-modulations confined to Hubbard band energies (Fig. 4a and b) supports this picture.

Despite the agreement of our experimental results with the existence of a QSL spinon FS, some mysteries remain. Two important open questions are the lack of experimental super-modulations at the $\mathbf{P}_i$ wavevectors and the origin of the experimental CS pattern at the K-point wavevectors. If the FS instability reflects a spinon density wave then we would expect to observe super-modulations at the primary $\mathbf{P}_i$ wavevectors. Therefore, the lack of an experimental signal at $\mathbf{P}_i$ suggests that the instability is due to other channels, such as a spinon pair density wave (PDW)[13,16] or a spinon spin density wave (SDW), implying that the super-modulations arise from higher-order processes. Such processes can be rationalized from either a scattering picture or a Landau formulation[40,41] (Supplementary Notes 3.2, 3.3) which lead to super-modulations at higher harmonic wavevectors (Supplementary Fig. 15). We note that $\mathbf{P}_i$ is also a possible second harmonic wavevector (Supplementary Fig. 15b), and so its experimental absence is likely a consequence of small structure factors at $\mathbf{P}_i$ (Supplementary Note 3.2). The observation of the CS at the K-point wavevectors, on the other hand, cannot be explained as a consequence of a spinon FS since K is not a harmonic of the $\mathbf{P}_i$ wavevectors. One possible explanation of the short-range CS is that our QSL phase is close to an antiferromagnetic ordered phase in the phase diagram, thus causing the CS to arise from short-range antiferromagnetic[42] fluctuations which are expected to be commensurate with the K-points (Supplementary Note 4).

**Conclusion**

In conclusion, we show strong evidence that single-layer 1T-$TaSe_2$ is a QSL with a spinon FS. Our observation of the Kondo effect in 1T-$TaSe_2$/1H-$TaSe_2$ heterostructures implies that SL 1T-$TaSe_2$ exhibits a triangular spin lattice, and our STS maps of 1T-$TaSe_2$ super-modulations directly reveal the effects of a spinon FS instability in this material. Our experimentally determined value of the spinon Fermi wavevector, $k_F = 0.380 \pm 0.005$ r.l.u., closely matches the theoretically predicted value of $k_F = 0.375$ r.l.u and supports the existence of a gapless QSL ground state. SL 1T-$TaSe_2$ thus provides an ideal platform to further investigate novel 2D QSL phenomena such as the response to magnetic scatterers[14,43,44] and electrostatic doping[6,7,45].

**Acknowledgments**

We thank Dung-Hai Lee, Joel E. Moore, and Michael Zaletel for helpful discussions.

**Funding:**

This research was supported by the VdW Heterostructure program (KCWF16) (STM/STS measurements) and the Advanced Light Source (sample growth and ARPES) funded by the Director, Office of Science, Office of Basic Energy Sciences, Materials Sciences and Engineering Division, of the US Department of Energy under Contract No. DE-AC02-05CH11231. Support was also provided by National Science Foundation award DMR-1807233 (surface treatment and topographic characterization) and DMR-1926004 (theoretical QPI analysis). The work at the Stanford Institute for Materials and Energy Sciences and Stanford University (ARPES measurements) was supported by the DOE Office of Basic Energy Sciences, Division of Material Science. The work at beamline 4-ID-D of the Advanced Photon Source, Argonne National Laboratory (X-ray absorption measurements) was supported by the DOE, Office of Science, Office of Basic Energy Sciences, under Contract No. DEAC02-06CH11357. P.A.L. acknowledges support by DOE Basic Energy Science award number DE-FG02-03ER46076 (theoretical QSL analysis). S. T. acknowledges the support by CPSF-CAS Joint Foundation for Excellent Postdoctoral Fellows. J.H. and C.H. acknowledge fellowship support from the NRF grant funded by the Korea government (MSIT) (No. 2018R1A2B6004538). H.-Z.T. acknowledges fellowship support from the Shenzhen Peacock Plan (Grant No. 827-000113, KQJSCX20170727100802505, KQTD2016053112042971).

**Author Contributions:**

W.R., Y.C., P.A.L., and M.F.C. initiated and conceived this project. W.R., Y.C., R.L., H.-Z.T., S.K., F.L., C.J., and A.A. carried out STM/STS measurements under the supervision of M.F.C. W.R., Y.C., F.W., P.A.L., and M.F.C. contributed to microscopy data analysis. S.T., J. H., and H.R. performed sample growth and ARPES measurements/analysis under the supervision of C.H., Z.-X.S., and S.-K.M. W.R., Y.C., R.L., and S.K. performed XMCD measurements with support from Y Choi. M.W. performed DFT+U calculations under the supervision of S.G.L. P.A.L. provided theoretical support. W.R., Y.C., and M.F.C. wrote the manuscript with the help from all authors. All authors contributed to the scientific discussion.

**Competing interests:**

The authors declare no competing interests.

**Data and materials availability:**

The data that support the findings of this study are available from the corresponding authors upon reasonable request.

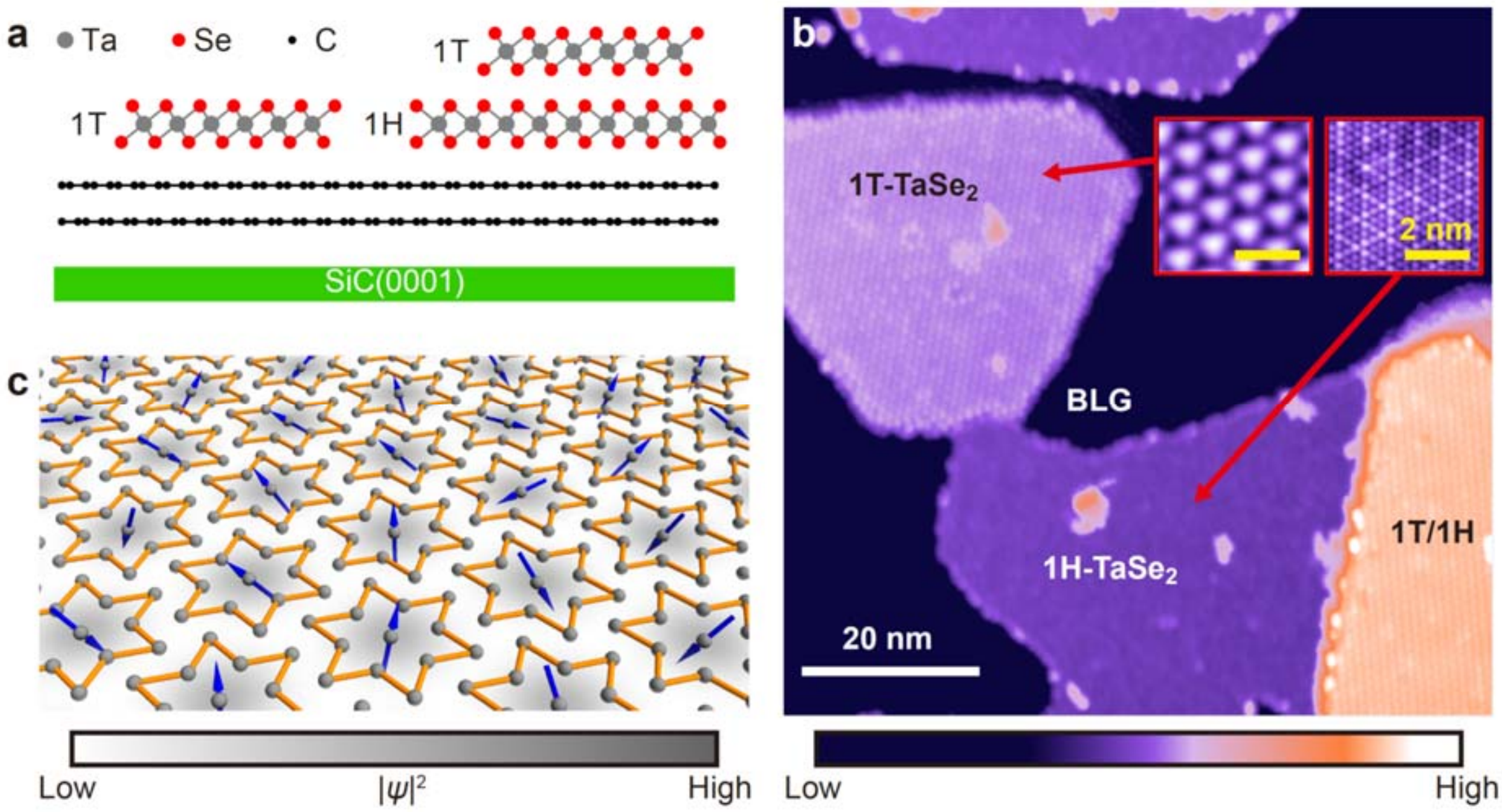

**Fig. 1. Structure of single-layer $TaSe_2$ and 1T/1H vertical heterostructures.**

**a**, Side-view of the crystal structures of single-layer 1T- and 1H-$TaSe_2$ as well as a 1T/1H vertical heterostructure on BLG-terminated SiC(0001). **b**, STM topographic image shows single-layer 1T-$TaSe_2$, single-layer 1H-$TaSe_2$, and a 1T/1H vertical heterostructure on BLG/SiC(0001) ($V_b$ = -1 V, $I_t$ = 5 pA). The insets are close-up images of the single-layer 1T and 1H islands (scanned at $V_b$ = -0.5 V, $I_t$ = 30 pA, and $V_b$ = 50 mV, $I_t$ = 1.3 nA, respectively) ($T$ = 5K). **c**, Schematic of the triangular spin lattice and star-of-David charge density wave pattern in 1T-$TaSe_2$. Each star consists of 13 Ta atoms and has a localized spin represented by a blue arrow at the star center. The wavefunction of the localized electrons is represented by gray shading.

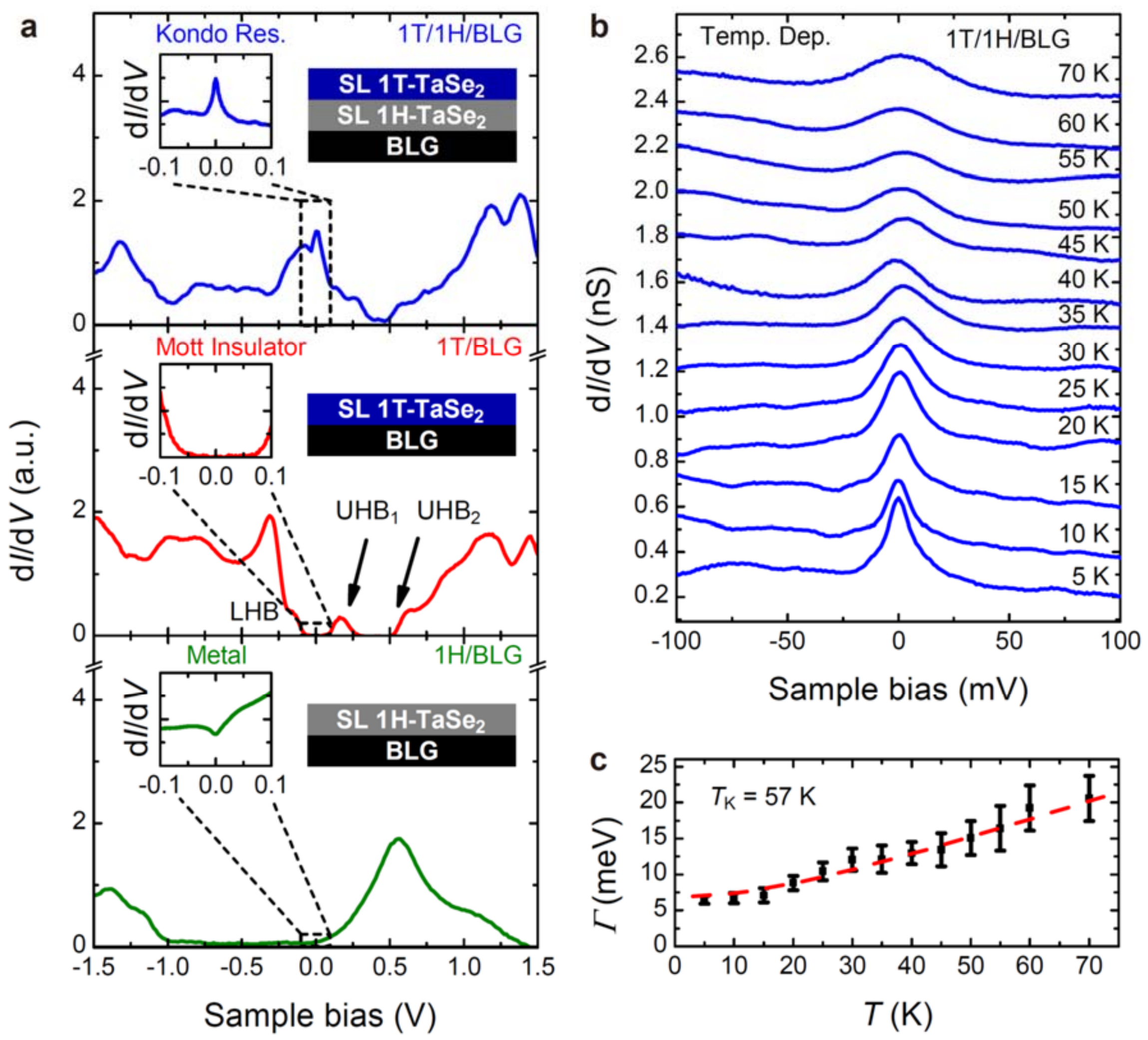

**Fig. 2. Kondo resonance observed in a 1T/1H $TaSe_2$ vertical heterostructure.**

**a**, Local electronic structure measured at $T = 5$ K via d$I$/d$V$ spectroscopy for single-layer 1H-$TaSe_2$ (green), single-layer 1T-$TaSe_2$ (red), and a 1T/1H $TaSe_2$ vertical heterostructure (blue) (1H: $V_b$ = -1.5 V, $I_t$ = 30 pA, $V_{mod}$ = 50 mV; 1T: $V_b$ = 1.5 V, $I_t$ = 40 pA, $V_{mod}$ = 20 mV; 1T/1H: $V_b$ = -1.5 V, $I_t$ = 30 pA, $V_{mod}$ = 20 mV). The insets show higher resolution d$I$/d$V$ spectra (1H: $V_b$ = 0.1 V, $I_t$ = 30 pA, $V_{mod}$ = 5 mV; 1T: $V_b$ = -0.25 V, $I_t$ = 30 pA, $V_{mod}$ = 2 mV; 1T/1H: $V_b$ = -0.1 V, $I_t$ = 30 pA, $V_{mod}$ = 1mV) (the Kondo peak in the inset is sharper and taller compared to the lower resolution measurement due to the use of a smaller wiggle voltage). **b**, Temperature dependence of the Kondo resonance peak observed in 1T/1H $TaSe_2$ vertical heterostructures for 5 K $\leqslant T \leqslant$ 70 K. The spectra are vertically offset for clarity ($V_b$ =

-0.1 V, $I_t$ = 30 pA, $V_{mod}$ = 1 mV). **c**, Temperature dependence of the intrinsic Kondo resonance width $\Gamma$ (black with error bar). The fit of Eq. (1) to the data (red dashed line) yields a Kondo temperature of $T_K$ = 57 K.

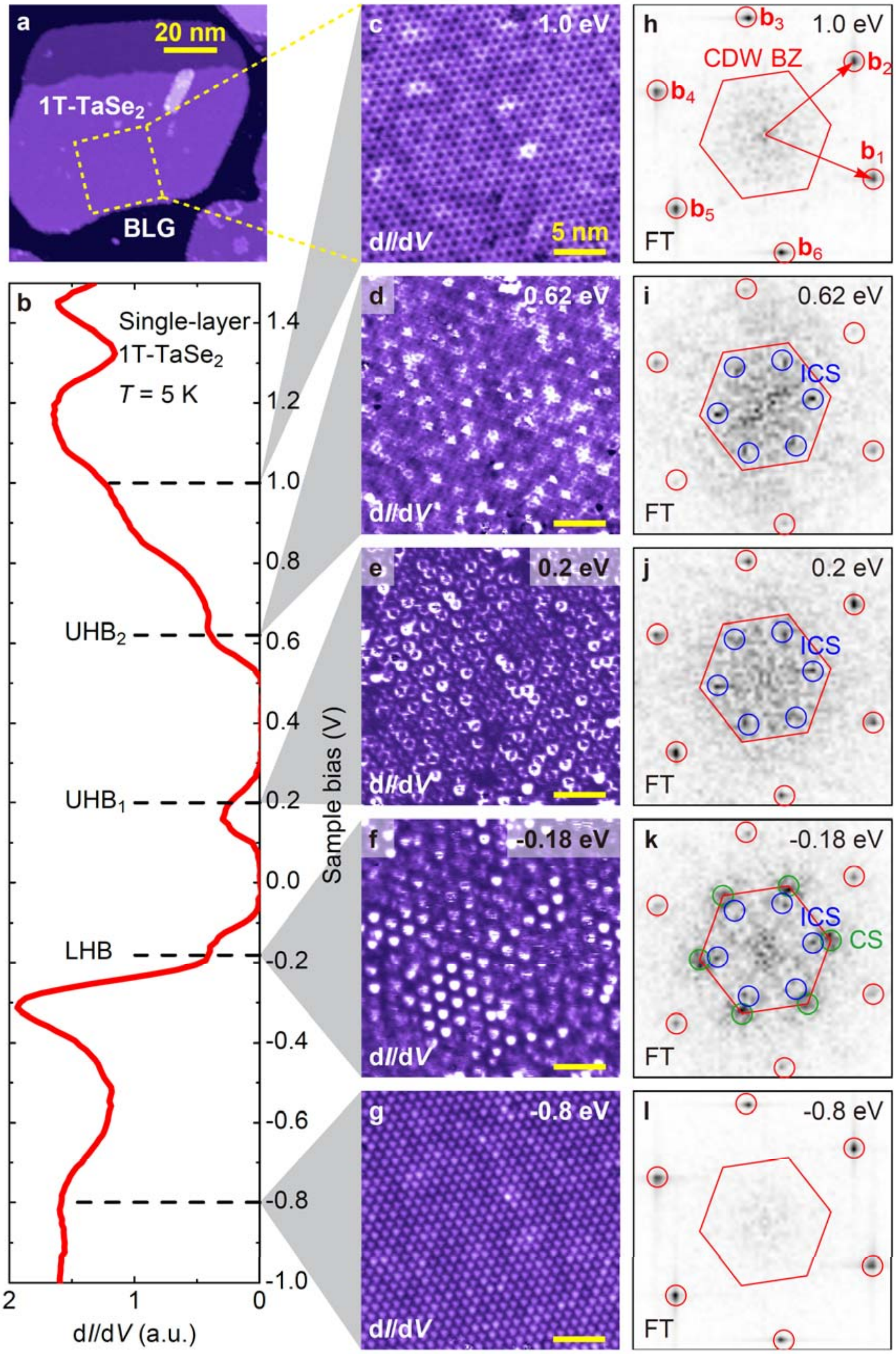

**Fig. 3. Super-modulations in single-layer 1T-TaSe₂ visualized by spectroscopic imaging.**

**a**, Large-scale topographic image of a single-layer 1T-$TaSe_2$ island ($V_b$ = -1 V, $I_t$ = 2 pA). **b**,

The d$I$/d$V$ spectrum of single-layer 1T-$TaSe_2$ ($V_b$ = 1.5 V, $I_t$ = 40 pA, $V_{mod}$ = 20 mV). **c-g**, Constant-height d$I$/d$V$ maps at different energies acquired in the area indicated by yellow dashed square in (A) ($I_t$ = 30 pA, $V_{mod}$ = 20 mV). **h-l**, Corresponding Fourier transforms (FTs) of the d$I$/d$V$ maps in **c-g**. FT peaks circled in red reflect the primary reciprocal lattice vectors of the star-of-David CDW and the red hexagon represents the CDW Brillouin zone (BZ). FT peaks circled in blue (**i-k**) reflect the incommensurate super-modulation (ICS) at 0.62 V, 0.2 V, and -0.18 V. FT peaks circled in green (**k**) reflect the commensurate super-modulation (CS) ($T$ = 5 K).

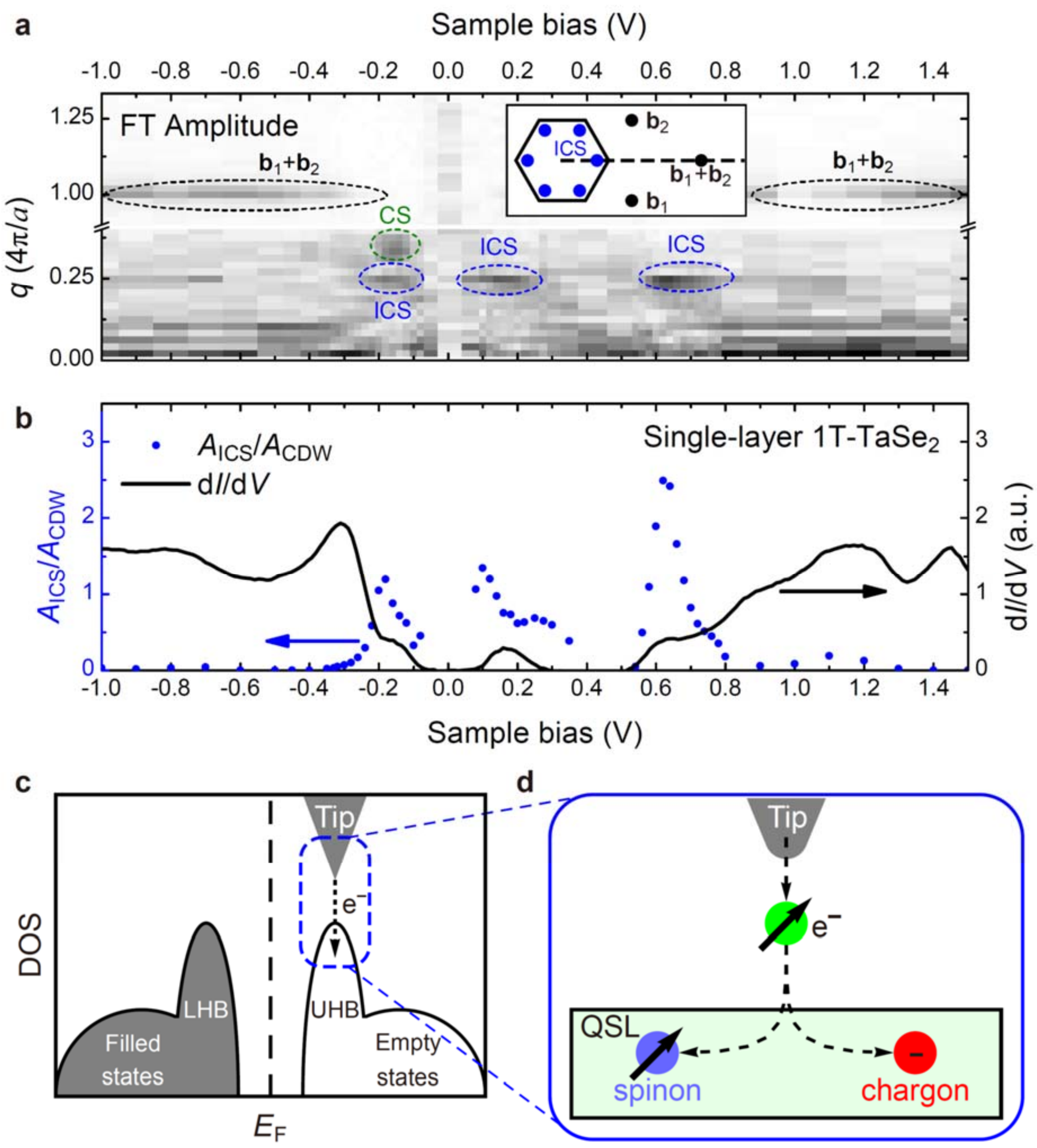

**Fig. 4. Energy dependence of super-modulations in single-layer 1T-TaSe$_2$.**

**a**, Plot of Fourier transform (FT) amplitude as a function of both the wavevector $\mathbf{q} = q(\mathbf{b}_1 + \mathbf{b}_2)$ ($q$ measured in units of $|\mathbf{b}_1 + \mathbf{b}_2| = 4\pi/a$) along the Γ-K direction ($y$-axis, indicated by black dashed line in the inset) and the sample bias voltage ($x$-axis). High FT amplitude (dark color) appears at $\mathbf{b}_1$+$\mathbf{b}_2$ (black dashed ovals), the ICS wavevector (blue dashed ovals), and the CS wavevector (green dashed oval) ($T$ = 5 K). **b**, Energy dependence of the ICS strength (blue dots) shows enhanced amplitude at Hubbard band energies. The ICS strength is defined as the FT amplitude at the ICS wavevector ($A_{ICS}$) divided by the FT amplitude at the CDW wavevector ($A_{CDW}$). The reference d*I*/d*V* spectrum is plotted in black ($T$ = 5 K). **c**, Schematic density of states of a Mott insulator. **d**, Cartoon of a tip-QSL tunnel junction. An electron injected into the strongly-correlated UHB of the QSL decays into a spinon and a chargon.

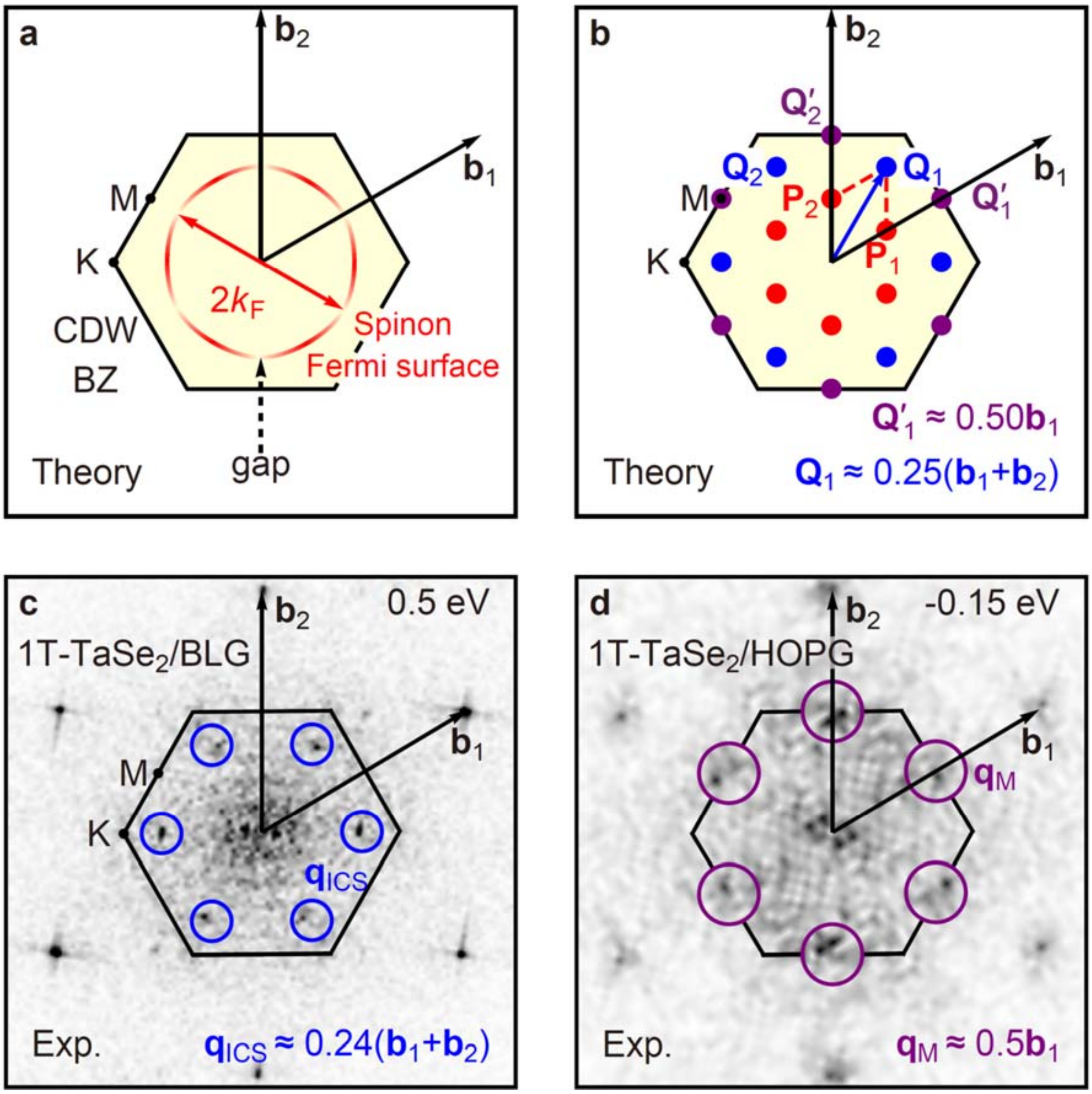

**Fig. 5. Super-modulation periodicities predicted from spinon Fermi surface (FS) compared with experiment.**

**a**, Schematic of spinon FS at half-filling (red) with partial gaps opening along the Γ-M directions. The black hexagon is the star-of-David CDW BZ and the black arrows ($\mathbf{b}_i$) are the primary reciprocal lattice vectors of the CDW BZ ($1 \leqslant i \leqslant 6$, only $\mathbf{b}_1$ and $\mathbf{b}_2$ are labeled). **b**, Spinon FS instability wavevectors $\mathbf{P}_i$ (red dots) and harmonics $\mathbf{Q}_i$ (blue dots) and $\mathbf{Q'}_i$ (purple dots) in the CDW BZ. **c**, Fourier transform (FT) of the STM image of 1T-$TaSe_2$/BLG at $V_b$ = 0.5 V (STM image shown in Supplementary Fig. 8a inset). Experimental ICS wavevectors $\mathbf{q}_{ICS}$ are circled in blue ($T$ = 5 K). **d**, FT of the STM image of 1T-$TaSe_2$/HOPG at $V_b$ = -0.15

V (STM image shown in Supplementary Fig. 12j). Experimental super-modulation wavevectors $\mathbf{q}_{\mathrm{M}}$ close to the M points are circled in purple ($T$ = 5 K).

**Supplementary Information:**

# Imaging spinon density modulation in a 2D quantum spin liquid

Wei Ruan[1,2,†], Yi Chen[1,2,†], Shujie Tang[3,4,5,6,7], Jinwoong Hwang[5,8], Hsin-Zon Tsai[1,10], Ryan Lee[1], Meng Wu[1,2], Hyejin Ryu[5,9], Salman Kahn[1], Franklin Liou[1], Caihong Jia[1,2,11], Andrew Aikawa[1], Choongyu Hwang[8], Feng Wang[1,2,12], Yongseong Choi[13], Steven G. Louie[1,2], Patrick A. Lee[14], Zhi-Xun Shen[3,4], Sung-Kwan Mo[5], Michael F. Crommie[1,2,12,*]

*[1]Department of Physics, University of California, Berkeley, California 94720, USA*

*[2]Materials Sciences Division, Lawrence Berkeley National Laboratory, Berkeley, California 94720, USA*

*[3]Stanford Institute for Materials and Energy Sciences, SLAC National Accelerator Laboratory and Stanford University, Menlo Park, California 94025, USA*

*[4]Geballe Laboratory for Advanced Materials, Departments of Physics and Applied Physics, Stanford University, Stanford, California 94305, USA*

*[5]Advanced Light Source, Lawrence Berkeley National Laboratory, Berkeley, California 94720, USA*

*[6]CAS Center for Excellence in Superconducting Electronics, Shanghai Institute of Microsystem and Information Technology, Chinese Academy of Sciences, Shanghai 200050, China*

*[7]School of Physical Science and Technology, Shanghai Tech University, Shanghai 200031, China*

*[8]Department of Physics, Pusan National University, Busan 46241, Korea*

*[9]Center for Spintronics, Korea Institute of Science and Technology, Seoul 02792, Korea*

*[10]International Collaborative Laboratory of 2D Materials for Optoelectronic Science & Technology of Ministry of Education, Engineering Technology Research Center for 2D Material Information Function Devices and Systems of Guangdong Province, Shenzhen University, Shenzhen 518060, China*

*[11]Henan Key Laboratory of Photovoltaic Materials and Laboratory of Low-dimensional Materials Science, School of Physics and Electronics, Henan University, Kaifeng 475004, China*

[12]*Kavli Energy Nano Sciences Institute at the University of California Berkeley and the Lawrence Berkeley National Laboratory, Berkeley, California 94720, USA*

[13]*Advanced Photon Source, Argonne National Laboratory, Argonne, Illinois 60439, USA*

[14]*Department of Physics, Massachusetts Institute of Technology, Cambridge MA 02139, USA*

*† These authors contributed equally to this work.*

**e-mail: crommie@berkeley.edu*

**Table of Contents**

**Materials and Methods**

**1. Sample growth and ARPES measurements**

Both sample growth and ARPES measurements were performed at the HERS endstation of Beamline 10.0.1, Advanced Light Source, Lawrence Berkeley National Laboratory. Single-layer $TaSe_2$ films were grown on both BLG/6H-SiC(001) and HOPG substrates in an ultrahigh-vacuum (UHV) MBE chamber under similar growth conditions as described elsewhere[1]. The substrate temperature was set at 660 °C, much higher than that for the growth of pure 1H-$TaSe_2$, to allow a simultaneous growth of single-layer 1T-$TaSe_2$ and single-layer 1H-$TaSe_2$, as well as vertical heterostructures composed of a 1T layer on top of a 1H layer. After growth, the samples were transferred in-situ into the analysis chamber with a base pressure $3\times10^{-11}$ Torr for ARPES measurements at 12 K. The photon energy was set at 50 eV with energy and angular resolution of 12 meV and 0.1°, respectively. The samples were then capped by Se capping layers with ~10 nm thickness for protection during transport through air to the UHV STM chamber.

**2. STM/STS measurements**

STM and STS measurements were performed in a low-temperature ultrahigh-vaccum STM system (CreaTec) at $T$ = 5 K (unless specified otherwise). Prior to measurements, the samples were annealed in UHV at ~200 °C for 3 hours to remove the Se capping layers and then immediately transferred in-situ into the STM stage sitting at $T$ = 5 K. Electrochemically etched tungsten tips were calibrated on a Au(111) surface before measurements. d$I$/d$V$ spectra were collected using standard lock-in techniques ($f$ = 401 Hz). d$I$/d$V$ mapping was performed in constant-height mode (i.e., with the feedback loop open).

**3. XMCD measurements**

XMCD measurements were performed at beamline 4-ID-D, Advanced Photon Source,

Argonne National Laboratory. Hard X-ray was used to probe Ta $L_2$ and $L_3$ edges at grazing incident angles to optimize the signal from single-layer 1T-$TaSe_2$ relative to background noise from the substrate. Absorption spectra were collected in fluorescence yield mode. SL 1T-$TaSe_2$/BLG samples were capped with Se layers during transport to the XMCD measurement chambers. The samples were mounted with the sample plane in the vertical direction. The X-ray beam size was adjusted to be 0.1 mm along the horizontal direction and 0.3 mm along the vertical direction.

**4. Thermal broadening simulation and thermally convolved Fano fit**

To determine the intrinsic temperature dependence of the Kondo resonance peak observed in the 1T/1H-$TaSe_2$ vertical heterostructure, we first performed a straightforward thermal broadening simulation:

$$\frac{\mathrm{d}I}{\mathrm{d}V}(eV) = \int_{-\infty}^{\infty} \rho(E)\big(-f'(E-eV,T)\big)\mathrm{d}E\,. \tag{1}$$

Here $\rho(E)$ is the zero-temperature density of states and $f(E,T) = \frac{1}{e^{E/k_\mathrm{B}T}+1}$ is the Fermi-Dirac distribution function at temperature $T$. In our simulation, $\rho(E)$ is approximated by the experimental d$I$/d$V$ curve at 5 K, where thermal broadening effects are negligible. The resulting thermally-broadened DOS (dashed red curves) shows that the temperature-induced broadening of the Kondo resonance peak cannot be accounted for by pure thermal broadening (Supplementary Fig. 4a).

We then performed a thermally convolved Fano fit using the following equation[2]:

$$\frac{\mathrm{d}I}{\mathrm{d}V}(eV) = \int_{-\infty}^{\infty} \left(a\frac{(\xi+q)^2}{1+\xi^2}+c\right)\big(-f'(E-eV,T)\big)\mathrm{d}E\,, \quad \xi = \frac{E-E_0}{\Gamma} \tag{2}$$

Here the first term in the integrand is a scaled Fano function plus a constant background, and the second term is the same thermal factor as in Eq. (1). In the Fano function, $E_0$ is the resonance energy, $\Gamma$ is the intrinsic Kondo resonance width, and $q$ is the ratio of the tunneling probability between the tip and the local moment over that between the tip and the itinerant electrons. This formula fits the experimental data well at all measured temperatures as seen in Supplementary Fig. 4b, thus allowing us to extract $\Gamma(T)$.

**5. Simulation of the spinon band structure**

In order to theoretically estimate the spinon Fermi wavevector $k_F$, we calculate the spinon band structure under a mean-field approximation where the emergent gauge fluctuations are neglected[3,4]. We consider a tight-binding Hamiltonian with uniform hopping amplitudes $t_s$ between nearest neighbors in the star-of-David triangular lattice (i.e., without emergent gauge flux):

$$H = -t_s \sum_{\langle ij \rangle, \sigma} \left( f_{i\sigma}^\dagger f_{j\sigma} + \text{h.c.} \right) - \mu \sum_{i\sigma} f_{i\sigma}^\dagger f_{i\sigma} , \tag{3}$$

where $\langle ij \rangle$ denotes the nearest-neighbor pair of site $i$ and site $j$, $\sigma$ is the spin index, and $f_{i\sigma}^\dagger$, $f_{i\sigma}$ are the spinon creation and annihilation operators. The chemical potential $\mu$ is included to ensure that the spinon band is half filled, a constraint required by the half-filling of electrons. Diagonalizing this tight-binding Hamiltonian results in the spinon dispersion

$$\varepsilon_{\mathbf{k}} = -2t_s(\cos \mathbf{k} \cdot \mathbf{a}_1 + \cos \mathbf{k} \cdot \mathbf{a}_2 + \cos \mathbf{k} \cdot \mathbf{a}_3) - \mu, \tag{4}$$

where $\mathbf{k}$ is the spinon momentum and the $\mathbf{a}_i$'s ($i$ = 1, 2, 3) are the nearest neighbor vectors in the triangular lattice with lattice constant $a$ (Supplementary Fig. 14a). The resulting spinon band structure is plotted in Supplementary Fig. 14b. The half-filling constraint yields a chemical potential of $\mu = 0.8346 t_s$. The spinon FS can be then obtained by solving the equation $\varepsilon_{\mathbf{k}} = 0$, yielding a slightly distorted circle that encloses exactly half the area of the whole CDW BZ (Supplementary Fig. 14c). In particular, the spinon Fermi wavevector along the Γ-M direction is $k_F \approx 0.375$ reciprocal lattice units (r.l.u.), rather close to the value of 0.371 estimated from a circular spinon FS assuming a half-filled isotropic spinon band.

## Supplementary Text

### Supplementary Note 1: Energy dependence of the star-of-David CDW pattern

The star-of-David CDW reconstruction persists at almost all energies from -1 eV to 1.5 eV, although the intra-star pattern can vary with energy (Supplementary Figs. 10 and 11) as previously reported[1]. For example, at $V_b$ = 0.2 V in the $UHB_1$ (Fig. 3e), each star-of-David cell manifests as flower-like pattern, i.e. the center of the cell is dark while the periphery exhibits bright "petals". This has been investigated and discussed in detail in Ref.[1].

### Supplementary Note 2: Kondo resonance

#### 2.1 Kondo resonance for 1T/1H heterostructure

The Kondo temperature $T_K$ = 57 ± 3 K obtained from our experiment characterizes the binding energy of the Kondo singlet formed in the 1T/1H vertical heterostructure. This allows us to estimate the Kondo coupling, $J_K$, between a localized spin in single-layer 1T-$TaSe_2$ and itinerant electrons in single-layer 1H-$TaSe_2$. The relation between $J_K$ and $T_K$ is[5,6]

$$T_K = D \exp\left(-\frac{1}{J_K \rho(0)}\right), \tag{5}$$

where $2D$ is the band width of single-layer 1H-$TaSe_2$, which is about 1 eV[7,8]. $\rho(0)$ is the Fermi-level density of states of the itinerant electrons in single-layer 1H-$TaSe_2$, which can be approximated by $\rho(0) = 1/(2D) = 1$ $eV^{-1}$ for a 2D electron gas. The Kondo coupling obtained from this treatment is $J_K \approx 0.2$ eV.

The observed zero-bias peak (ZBP) in the 1T/1H heterostructure is not due to strain effects but rather arises from the Kondo effect. The observed topographic and spectroscopic features of our van der Waals coupled samples are very different from those reported for a sister material, bulk 1T-$TaS_2$, which exhibit strain effects induced by special sample treatment[9-11]. We have also carefully measured the star-of-David CDW periodicity of the 1T layer in both the 1T/1H heterostructure (12.29 ± 0.38 Å) and bulk 1T-$TaSe_2$ (12.52 ± 0.31 Å) which does not suffer from lattice mismatch. Both values are within the measurement error from each other.

### 2.2 Kondo lattice behavior in the 1T/1H heterostructure

Technically, a Kondo lattice is formed in the 1T/1H vertical heterostructure because there is a triangular lattice of local moments in the single-layer Mott insulator 1T-$TaSe_2$ coupled to the metallic 1H-$TaSe_2$ underneath. In principle, coherent Kondo lattice behavior can emerge, including renormalized heavy band formation[5] and direct exchange or Ruderman-Kittel-Kasuya-Yosida (RKKY) mediated antiferromagnetism[12]. However, our observations are consistent with single Kondo impurity behavior rather than coherent Kondo lattice behavior. The likely reason for this is that coherent Kondo lattice behavior occurs below a coherence temperature[6] $T^* \ll T_K$. In our system the base temperature (5 K) of the measurements is likely not low enough for the observation of coherent behavior since $T_K$ = 57 K.

We can estimate the value of in-plane superexchange $J$ between local moments in the single-layer Mott insulator 1T-$TaSe_2$ based on the measured Kondo temperature $T_K$ in the 1T/1H heterostructure by considering the Doniach phase diagram for a Kondo lattice[5,12]. Here the ground state is determined by a competition between the Kondo temperature $T_K$ (which characterizes the Kondo singlet energy scale) and the RKKY exchange coupling $J_{RKKY}$ (which characterizes the antiferromagnetic ordering energy scale). $T_K$ has to be larger than $J_{RKKY}$ for the system to fall into the Kondo phase. For our QSL/metal system, $J$ plays a similar role to $J_{RKKY}$ in the sense that they both represent antiferromagnetic exchange between local spins, and so $J$ must be smaller than $T_K$ in order to see the Kondo resonance. Given our measured Kondo temperature $T_K \sim 60$ K, we estimate an upper limit of $J$ to be about 5 meV. However, we emphasize that, unlike the competition between the Néel state and the Kondo singlet, the competition between a QSL state and the Kondo singlet is a new problem that has not been studied. Taking into consideration that the QSL state and the Kondo singlet both favor strong fluctuations of the local spins rather than static ordering, the upper limit of $J$ imposed by $T_K$ could be relaxed (i.e., $J$ might be larger than $T_K$).

**Supplementary Note 3: The incommensurate super-modulation (ICS)**

**3.1 Alternative explanations for the ICS**

The observed super-modulations in SL 1T-$TaSe_2$/BLG are not due to strain effects. The observed topographic and spectroscopic features of our van der Waals coupled samples are very different from those reported for a sister material, bulk 1T-$TaS_2$, which exhibit strain effects induced by special sample treatment[9-11]. The measured star-of-David CDW periodicity of 1T single-layer on BLG is 12.52 ± 0.39 Å, which is indistinguishable from the bulk 1T-$TaSe_2$ value of 12.52 ± 0.31 Å.

**3.2 Phenomenological Landau theory of composite spinon density wave (sDW) orders and spinon PDW/SDW orders**

In this section, we explain how super-modulations observable by STM at higher harmonic wavevectors, $\mathbf{q}_1 \pm \mathbf{q}_2$, might form in the presence of hidden fundamental orders at wavevectors $\pm\mathbf{q}_1$ and $\pm\mathbf{q}_2$ even if they are fluctuating and short-ranged[13,14]. The two candidates we consider for the fundamental orders are spinon pair density wave (sPDW) and spinon spin density wave (sSDW) order. Here we extend the previous phenomenological Landau theory from the 1D and square lattice[13-15] to the triangular lattice. The difference is that on the triangular lattice in our case, there are six independent complex order parameters (e.g., sPDW) at wavevectors $\pm\mathbf{P}_i$ ($i$ = 1, 2, 3 as shown in Fig. 5 and Supplementary Fig. 15, where $\mathbf{P}_4 = -\mathbf{P}_1, \mathbf{P}_5 = -\mathbf{P}_2, \mathbf{P}_6 = -\mathbf{P}_3$) or three independent complex order parameters (e.g., colinear sSDW) at wavevectors $\pm\mathbf{P}_i$ (in this case the order parameter is real in space and therefore the order parameters at $+\mathbf{P}_i$ and $-\mathbf{P}_i$ are related by complex conjugation). The case of non-colinear SDW such as the coplanar 120° order will be discussed in Note 4. We shall find that much of the conclusions reached in Ref.[14] remains the same, namely, the induced composite order at $\mathbf{q}_1 \pm \mathbf{q}_2$ is composite spinon density wave (sDW). The composite sDW order will induce electron density super-modulations observable by STM at higher harmonic wavevectors (Supplementary Note 3.4).

In the following we focus our attention on two sets of fundamental wavevectors $\mathbf{q}_1 = \mathbf{P}_1$ and $\mathbf{q}_2 = \mathbf{P}_2$. Contributions from the remaining wavevectors $\pm\mathbf{P}_3$ can be easily added due to the 6-fold in-plane rotational symmetry.

To understand how sDW order develops in a Landau theory framework, we start with the simplest case where the Landau free energy density in a system with translational, rotational, and gauge symmetries (which conserves momentum, angular momentum, and particle number, respectively)[13-15] can be written (up to fourth order) as

$$F_{\mathrm{sDW}}^{0}(\mathbf{q}) = \frac{\kappa_{\mathbf{q}}}{2}\left|\nabla\rho_{\mathbf{q}}\right|^2 + \frac{m_{\mathbf{q}}}{2}\left|\rho_{\mathbf{q}}\right|^2 + \frac{\gamma_{\mathbf{q}}}{4}\left|\rho_{\mathbf{q}}\right|^4, \quad (6)$$

where $\rho_{\mathbf{q}}$ is the sDW order parameter with wavevector $\mathbf{q}$, $\kappa_{\mathbf{q}}$ and $\gamma_{\mathbf{q}}$ are positive to ensure that the free energy does not go to negative infinity, and the sign of $m_{\mathbf{q}}$ determines the ground state. With several $\mathbf{q}$'s, additional fourth order terms are allowed, but they are not essential for our discussion and shall not be considered further. The normal phase without sDW order can be achieved with a positive mass $m_{\mathbf{q}} > 0$, which leads to a free energy minimum at $\rho_{\mathbf{q}} = 0$. In contrast, the sDW phase can be achieved for negative mass ($m_{\mathbf{q}} < 0$), which leads to free energy minima at nonzero values of $\left|\rho_{\mathbf{q}}\right| = \sqrt{-m_{\mathbf{q}}/\gamma_{\mathbf{q}}}$.

We now discuss how composite sDW order at higher harmonic wavevector, $\mathbf{q}_1 + \mathbf{q}_2$, can form from sPDW order at wavevectors $\pm\mathbf{q}_1$ and $\pm\mathbf{q}_2$ (i.e. in the presence of nonzero order parameters $\Delta_{\pm\mathbf{q}_1} \neq 0$ and $\Delta_{\pm\mathbf{q}_2} \neq 0$). Unlike the case above, composite sDW order $\rho_{\mathbf{q}_1+\mathbf{q}_2}$ at $\mathbf{q}_1 + \mathbf{q}_2$ can be achieved regardless of the sign of the mass $m_{\mathbf{q}_1+\mathbf{q}_2}$ so long as the composite sDW order is coupled to the sPDW order at $\pm\mathbf{q}_1$ and $\pm\mathbf{q}_2$ (i.e., $\Delta_{\pm\mathbf{q}_1}$ and $\Delta_{\pm\mathbf{q}_2}$). In this case the total Landau free energy density that conserves momentum and particle number is[13,14]

$$F = F^0_{\mathrm{sPDW}} + F^0_{\mathrm{sDW}}(\mathbf{q}_1 + \mathbf{q}_2)$$

$$+\left[\lambda_{\mathbf{q}_1,\mathbf{q}_2}\rho_{\mathbf{q}_1+\mathbf{q}_2}\left(\Delta^*_{\mathbf{q}_1}\Delta_{-\mathbf{q}_2} + \Delta^*_{\mathbf{q}_2}\Delta_{-\mathbf{q}_1}\right) + \mathrm{h.c.}\right] \quad (7)$$

where $F^0_{\mathrm{sPDW}}$ is the free energy density for sPDW order at $\pm\mathbf{q}_1$ and $\pm\mathbf{q}_2$, and the interaction term is kept to lowest order. By varying $\rho_{\mathbf{q}_1+\mathbf{q}_2}$ while keeping $\Delta_{\pm\mathbf{q}_1}$ and $\Delta_{\pm\mathbf{q}_2}$ at fixed nonzero values, the above free energy density can be minimized at a nonzero value of $\rho_{\mathbf{q}_1+\mathbf{q}_2}$, indicating the formation of composite sDW order at the higher harmonic wavevector $\mathbf{q}_1 + \mathbf{q}_2 = \mathbf{P}_1 + \mathbf{P}_2$ and similarly at $\mathbf{P}_1 - \mathbf{P}_2$ if we take $\mathbf{q}_2 = -\mathbf{P}_2$ instead.

The relative strength of the composite sDW at $\mathbf{P}_1 + \mathbf{P}_2$ (for $\mathbf{q}_1 = \mathbf{P}_1$, $\mathbf{q}_2 = \mathbf{P}_2$) and $\mathbf{P}_1 - \mathbf{P}_2$ (for $\mathbf{q}_1 = \mathbf{P}_1$, $\mathbf{q}_2 = -\mathbf{P}_2$) is determined by the two different Landau coupling constants ($\lambda_{\mathbf{P}_1,\mathbf{P}_2}$ and $\lambda_{\mathbf{P}_1,-\mathbf{P}_2}$) and can be quite different. This may explain the absence of experimental charge density modulation at $\mathbf{P}_1 - \mathbf{P}_2$ (which happens to coincide with the fundamental wavevectors).

Similar arguments support the formation of composite sDW order at higher harmonic wavevector, $\mathbf{q}_1 + \mathbf{q}_2$, in the presence of sSDW order at $\pm\mathbf{q}_1$ and $\pm\mathbf{q}_2$ (i.e. due to the existence of nonzero order parameters $\mathbf{S}_{\pm\mathbf{q}_1} \neq 0$ and $\mathbf{S}_{\pm\mathbf{q}_2} \neq 0$). Here the total Landau free energy density that conserves momentum, particle number, and spin is[15]

$$F = F^0_{\mathrm{sSDW}} + F^0_{\mathrm{sDW}}(\mathbf{q}_1 + \mathbf{q}_2)$$

$$+\lambda_{\mathbf{q}_1,\mathbf{q}_2}\rho_{\mathbf{q}_1+\mathbf{q}_2}\left(\mathbf{S}_{-\mathbf{q}_1}\cdot\mathbf{S}_{-\mathbf{q}_2} + \mathbf{S}^*_{\mathbf{q}_1}\cdot\mathbf{S}^*_{\mathbf{q}_2}\right) + \mathrm{h.c.}, \quad (8)$$

where $F^0_{\mathrm{sSDW}}$ is the free energy density for sSDW order and the interaction term is kept to lowest order. By varying $\rho_{\mathbf{q}_1+\mathbf{q}_2}$ while keeping $\mathbf{S}_{-\mathbf{q}_1}$ and $\mathbf{S}_{-\mathbf{q}_2}$ at fixed nonzero values, this free energy density can be minimized at a nonzero value of $\rho_{\mathbf{q}_1+\mathbf{q}_2}$, indicating the formation of composite sDW order at the higher harmonic wavevector $\mathbf{q}_1 + \mathbf{q}_2$. For real order parameters such as colinear sSDW order, the interaction term generally does not vanish since $\mathbf{S}_{-\mathbf{q}} = \mathbf{S}_{\mathbf{q}}^*$, so that composite sDW is always produced.

To summarize, according to Landau theory for the triangular lattice, fundamental orders such as sPDW or sSDW order at primary wavevectors $\mathbf{P}_i$ can induce composite sDW order at high harmonic wavevectors $\mathbf{P}_i + \mathbf{P}_j$ ($1 \leqslant i, j \leqslant 6$). The strengths of induced composite sDW order at different wavevectors, which are determined by the coupling constant in Landau theory, can be quite different.

### 3.3 A scattering picture for the development of super-modulations at higher harmonics

Another way to rationalize the formation of super-modulations observable by STM at higher harmonic wavevectors is through a single-particle scattering picture in the presence of either spinon pair density wave (sPDW) order or spinon spin density wave (sSDW) order.

We first consider the case of sPDW order at wavevectors $\mathbf{q}_1$ and $-\mathbf{q}_2$. In this scenario spinon Cooper pairs are expected to exist that simultaneously involve center-of-mass momenta $\mathbf{q}_1$ and $-\mathbf{q}_2$. The center-of-mass wavefunction of such Cooper pairs can be approximated by a superposition of plane waves

$$\psi_{\mathrm{pair}}(\mathbf{r}) \propto c_1 \mathrm{e}^{\mathrm{i}\mathbf{q}_1\cdot\mathbf{r}} + c_2 \mathrm{e}^{-\mathrm{i}\mathbf{q}_2\cdot\mathbf{r}}. \tag{9}$$

The spinon Cooper pair density of such pairs, $n_{\mathrm{pair}}(\mathbf{r})$, is proportional to $|\psi_{\mathrm{pair}}|^2$:

$$n_{\mathrm{pair}}(\mathbf{r}) = |\psi_{\mathrm{pair}}|^2 \propto |c_1|^2 + |c_2|^2 + 2\,\mathrm{Re}\left(c_1 c_2^* \mathrm{e}^{\mathrm{i}(\mathbf{q}_1+\mathbf{q}_2)\cdot\mathbf{r}}\right), \tag{10}$$

which exhibits a spatial periodicity at the wavevector $\mathbf{q}_1 + \mathbf{q}_2$. Such spinon Cooper pairs can interact with single spinons via spinon-spinon interactions, thus generating a scattering potential that follows the spatial profile of $n_{\mathrm{pair}}(\mathbf{r})$ (such spinon-spinon interactions could be mediated by, for example, the same emergent U(1) gauge bosons in the gapless QSL state that are also responsible for spinon pairing[16]). A single spinon scattering off such a periodic potential will form a density wave pattern at wavevector $\mathbf{q}_1 + \mathbf{q}_2$, much like electrons scattering off a periodic atomic lattice. When recombined with chargons, the resulting spatially periodic spinon pattern will cause a super-modulation at the higher harmonic wavevector $\mathbf{q}_1 + \mathbf{q}_2$ in the quasiparticle channel that can be detected by STM (Supplementary Note 3.4).

For the alternative case of sSDW order at wavevectors $\mathbf{q}_1$ and $\mathbf{q}_2$, the total spin density takes the form

$$\mathbf{S}(\mathbf{r}) \propto \mathbf{s}_1 \cos(\mathbf{q}_1 \cdot \mathbf{r} + \phi_1) + \mathbf{s}_2 \cos(\mathbf{q}_2 \cdot \mathbf{r} + \phi_2). \tag{11}$$

This spin density can, in principle, induce a spin-dependent scattering potential $\mathbf{V}_0(\mathbf{r}) \propto \mathbf{S}(\mathbf{r})$ for single spinons described by a two-component spinor wavefunction. For example, the scattering potential felt by a spin-up spinon can be different from the potential felt by a spin-down spinon depending on the local sSDW environment, and thus the potential should be proportional to $S_z$ (similar arguments hold for both $S_x$ and $S_y$). In order to obtain a spinon density super-modulation, a scalar (spin-independent) scattering potential is needed. The lowest-order scalar potential in the presence of the sSDW order (Eq. (11)) is achieved by taking the square of the potential $\mathbf{V}_0(\mathbf{r})$ corresponding to a second-order process, i.e.

$$\begin{aligned} V(\mathbf{r}) &\propto |\mathbf{V}_0(\mathbf{r})|^2 \\ &\propto \mathbf{s}_1^2 \frac{1 + \cos 2(\mathbf{q}_1 \cdot \mathbf{r} + \phi_1)}{2} + \mathbf{s}_2^2 \frac{1 + \cos 2(\mathbf{q}_2 \cdot \mathbf{r} + \phi_2)}{2} \\ &+ \mathbf{s}_1 \cdot \mathbf{s}_2 \left[\cos\left((\mathbf{q}_1 + \mathbf{q}_2) \cdot \mathbf{r} + \phi_1 + \phi_2\right) + \cos\left((\mathbf{q}_1 - \mathbf{q}_2) \cdot \mathbf{r} + \phi_1 - \phi_2\right)\right]. \end{aligned} \tag{12}$$

This potential exhibits modulations at the higher harmonic wavevector $\mathbf{q}_1 + \mathbf{q}_2$ (also $2\mathbf{q}_1$, $2\mathbf{q}_2$ and $\mathbf{q}_1 - \mathbf{q}_2$), thus leading to spatially modulated spinon density at $\mathbf{q}_1 + \mathbf{q}_2$.

### 3.4 Spinon and chargon spectral function convolution

In this section we use a spectral function convolution process to explain how spinon density modulations can be imaged by STM in the electron channel.

In a weakly coupled model of nearly free spinons and chargons, the electron Green's function to the lowest-order approximation is a convolution of the free spinon and chargon Green's functions in momentum and energy[17-19]. The STM d$I$/d$V$ signal is proportional to the electron local density of states (LDOS) or spectral function $\rho(E, \mathbf{r})$. The electron spectral function can be written as an integral over the chargon energy $E_\mathrm{c}$ (prefactors such as the form factor are neglected)[18]:

$$\rho(E, \mathbf{r}) = \int_{E_\mathrm{c}>0} \big(1 - f(E - E_\mathrm{c})\big) \rho_\mathrm{s}(E - E_\mathrm{c}, \mathbf{r}) \rho_\mathrm{c}(E_\mathrm{c}, \mathbf{r}) \mathrm{d}E_\mathrm{c}$$

$$+ \int_{E_\mathrm{c}<0} f(E - E_\mathrm{c}) \rho_\mathrm{s}(E - E_\mathrm{c}, \mathbf{r}) \rho_\mathrm{c}(E_\mathrm{c}, \mathbf{r}) \mathrm{d}E_\mathrm{c}. \quad (13)$$

Here $\rho$, $\rho_\mathrm{s}$, and $\rho_\mathrm{c}$ are the electron, spinon, and chargon LDOS, respectively, and $f(E)$ is the Fermi-Dirac distribution function. In the following we take the zero-temperature limit $f(E) = \theta(-E)$ where $\theta$ is the Heaviside step-function. For our purpose it is more useful to consider reciprocal space where the Fourier-transform LDOS (FT-LDOS) of electrons with energy $E$ and wavevector $\mathbf{q}$ is an integral of the chargon energy $E_\mathrm{c}$ and wavevector $\mathbf{q}_\mathrm{c}$:

$$\rho(E, \mathbf{q}) = \int_{\substack{E_\mathrm{c}>0 \\ E-E_\mathrm{c}>0}} \rho_\mathrm{s}(E - E_\mathrm{c}, \mathbf{q} - \mathbf{q}_\mathrm{c}) \rho_\mathrm{c}(E_\mathrm{c}, \mathbf{q}_\mathrm{c}) \mathrm{d}E_\mathrm{c} \mathrm{d}\mathbf{q}_\mathrm{c}$$

$$+ \int_{\substack{E_\mathrm{c}<0 \\ E-E_\mathrm{c}<0}} \rho_\mathrm{s}(E - E_\mathrm{c}, \mathbf{q} - \mathbf{q}_\mathrm{c}) \rho_\mathrm{c}(E_\mathrm{c}, \mathbf{q}_\mathrm{c}) \mathrm{d}E_\mathrm{c} \mathrm{d}\mathbf{q}_\mathrm{c}. \quad (14)$$

As explained in Supplementary Notes 3.2 and 3.3, spinon density modulation can be induced at the theoretical wavevectors $\mathbf{Q}_i$ (Supplementary Fig. 15) which means that the spinon FT-LDOS has nonzero component $\rho_\mathrm{s}(E_\mathrm{s}, \mathbf{q} = \mathbf{Q}_i)$. As seen from Eq. (14), in general the electron LDOS should also have nonzero components at $\mathbf{q} = \mathbf{Q}_i$ which can be visualized by STM. As an example to show this idea, we assume that the spinon band spans the energy range $E_1 \leq E_s \leq E_2$ ($E_1 < 0$, $E_2 > 0$) and that the leading spinon density modulations are at $\mathbf{Q}_i$. The electron FT-LDOS in the LHB ($E < 0$) at $\mathbf{q} = \mathbf{Q}_i$ is the following (similar for the

UHB):

$$\rho(E,\mathbf{Q}_i) = \int_{\substack{E_c<0 \\ E-E_c<0}} \rho_s(E-E_c,\mathbf{Q}_i)\rho_c(E_c,\mathbf{q}_c=0)\mathrm{d}E_c$$

$$= \int_{\substack{E_1<E_s<0 \\ E-E_s<0}} \rho_s(E_s,\mathbf{Q}_i)\rho_c(E-E_s,\mathbf{q}_c=0)\mathrm{d}E_s\,, \tag{15}$$

which is a sum of the spinon $\mathbf{Q}_i$ FT-LDOS over spinon energy $E_s$ weighted by the energy-dependent chargon $\mathbf{q}=0$ FT-LDOS (which is the spatially averaged LDOS proportional to the chargon total DOS $\rho_c(E)$). For the spinon density wave state the following integral

$$\int_{E_1<E_s<0} \rho_s(E_s,\mathbf{Q}_i)\mathrm{d}E_s$$

represents the $\mathbf{Q}_i$ density modulation of all occupied spinons and is thus nonzero. Therefore, $\rho(E,\mathbf{Q}_i)$ in Eq. (15) is also generally nonzero for all the energies $E$ in the LHB, as long as the chargon DOS $\rho_c(E_c)$ is a slowly varying function in the small energy window $E_c \in [E, E-E_1]$. This is achieved as long as the chargon bandwidth is larger than or comparable to the spinon bandwidth which is proportional to the spin-exchange $J$. Our rough estimate of $J$ in Supplementary Note 2.2 is smaller than the Hubbard bandwidth and is therefore consistent with this condition. Based on the above analysis, the convolution process explains the fact that our observed super-modulations exist over the entire Hubbard band energy range beyond the characteristic spinon energy scale.

To see how the observed ICS modulation amplitude can be comparable to that of the star-of-David CDW pattern, we define the relative strength of the ICS over the star-of-David CDW as $r(E) = |\rho(E,\mathbf{Q}_i)|/|\rho(E,\mathbf{b}_i)|$ where $\mathbf{b}_i$ are the star-of-David reciprocal lattice vectors (Fig. 3). The experimental value of $r(E)$ can take values of around 1 ~ 3 as seen in Fig. 4b. Conceptually, the ICS mainly reflects the spatial wavefunction of delocalized spinons, while the star-of-David CDW reflects the spatial wavefunction of localized chargons subject to various Coulomb interactions (e.g., chargon-chargon, chargon-phonon). To model this we assume a nonzero chargon FT-LDOS at $\mathbf{b}_i$: $\rho_c(E_c, \mathbf{b}_i) = \alpha_c(E_c, \mathbf{b}_i)\rho_c(E_c, 0)$ where $\alpha_c(E_c, \mathbf{b}_i)$ represents the relative strength of the chargon density modulation at $\mathbf{b}_i$ compared to the spatially averaged chargon LDOS $\rho_c(E_c, 0)$. Similarly we can write the spinon FT-LDOS at $\mathbf{Q}_i$ as $\rho_s(E_s, \mathbf{Q}_i) = \alpha_s(E_s, \mathbf{Q}_i)\rho_s(E_s, 0)$ where $\alpha_s(E_s, \mathbf{Q}_i)$ represents the relative strength of the spinon density modulation at $\mathbf{Q}_i$ to the spatially averaged spinon LDOS $\rho_s(E_s, 0)$. Then we have the ratio $r(E)$ in the LHB as follows (similar for the UHB):

$$r(E) = \left|\frac{\rho(E, \mathbf{Q}_i)}{\rho(E, \mathbf{b}_i)}\right| = \left|\frac{\int_{\substack{E_1<E_s<0 \\ E-E_s<0}} \alpha_s(E_s, \mathbf{Q}_i)\rho_s(E_s, 0)\rho_c(E - E_s, 0)\mathrm{d}E_s}{\int_{\substack{E_1<E_s<0 \\ E-E_s<0}} \alpha_c(E - E_s, \mathbf{b}_i)\rho_s(E_s, 0)\rho_c(E - E_s, 0)\mathrm{d}E_s}\right|. \tag{16}$$

From Eq. (16) there are at least two reasons for the experimental observation of the $r(E)$ value being around 1 ~ 3 (Fig. 4b). First, $|r(E)|$ is roughly determined by the ratio of $\alpha_s$ over $\alpha_c$ because the numerator and the denominator are integrals of $\alpha_s$ and $\alpha_c$ weighted by the same factor $\rho_s(E_s, 0)\rho_c(E - E_s, 0)$, respectively. $\alpha_s$ and $\alpha_c$ are two relatively independent parameters that can in principle take the ratio of 1 ~ 3 ($\alpha_s$ in this case does not have to be much smaller than $\alpha_c$). Second, the intra-cell structure of the star-of-David CDW exhibits a strong energy dependence[1] (see also Fig. 3 and Supplementary Figs. 10 and 11) indicating that the phase of the complex quantity $\alpha_c(E_c, \mathbf{b}_i)$ varies strongly as a function of the chargon energy. Therefore, the integral in the denominator can be influenced by destructive interference, resulting in a small absolute value of $|\rho(E, \mathbf{b}_i)|$.

We have also performed numerical simulation of the spectral function convolution to further illustrate these ideas. To simulate the spinon spectral function we start from the free spinon band dispersion $\varepsilon_{\mathbf{k}}$ shown in Eq. (4). The spinon density wave state at wavevectors $\mathbf{Q}_i$ ($i = 1,2, \dots, 6$) is modeled phenomenologically by considering a periodic scattering potential at wavevectors $\mathbf{Q}_i$ for the spinons. This results in a momentum-dependent Hamiltonian for the spinons as follows:

$$H_s(\boldsymbol{k}) = \begin{bmatrix} \varepsilon_{\mathbf{k}} & \Delta_{\mathbf{k},\mathbf{k}+\mathbf{Q}_1} & \cdots & \Delta_{\mathbf{k},\mathbf{k}+\mathbf{Q}_6} \\ \Delta^*_{\mathbf{k},\mathbf{k}+\mathbf{Q}_1} & \varepsilon_{\mathbf{k}+\mathbf{Q}_1} & 0 & 0 \\ \vdots & 0 & \ddots & \vdots \\ \Delta^*_{\mathbf{k},\mathbf{k}+\mathbf{Q}_6} & 0 & \cdots & \varepsilon_{\mathbf{k}+\mathbf{Q}_6} \end{bmatrix}, \tag{17}$$

where $\Delta_{\mathbf{k},\mathbf{k}+\mathbf{Q}_i} \equiv \Delta_0$ represents the Fourier component of the spinon scattering potential and is taken as a constant for simplicity. The above Hamiltonian can be diagonalized as

$$H_s(\mathbf{k}) \begin{bmatrix} c_{n\mathbf{k}} \\ c_{n\mathbf{k}+\mathbf{Q}_1} \\ \vdots \\ c_{n\mathbf{k}+\mathbf{Q}_6} \end{bmatrix} = E_n(\mathbf{k}) \begin{bmatrix} c_{n\mathbf{k}} \\ c_{n\mathbf{k}+\mathbf{Q}_1} \\ \vdots \\ c_{n\mathbf{k}+\mathbf{Q}_6} \end{bmatrix}, \quad n = 1,2,\dots,7, \tag{18}$$

with the spatial spinon wavefunction

$$\psi_{n\mathbf{k}}(\mathbf{r}) = c_{n\mathbf{k}} \mathrm{e}^{\mathrm{i}\mathbf{k}\cdot\mathbf{r}} + \sum_{j=1}^{6} c_{n\mathbf{k}+\mathbf{Q}_j} \mathrm{e}^{\mathrm{i}(\mathbf{k}+\mathbf{Q}_j)\cdot\mathbf{r}}. \tag{19}$$

Then the spinon LDOS $\rho_s(E, \mathbf{r})$ is

$$\rho_s(E, \mathbf{r}) = \sum_{n\mathbf{k}} |\psi_{n\mathbf{k}}(\mathbf{r})|^2 \delta\big(E - E_n(\mathbf{k})\big), \tag{20}$$

where the Dirac-$\delta$ function is replaced by a Gaussian smearing function in practice. For the chargons that form a 2D Bose gas, we assume a constant chargon DOS for simplicity:

$$\rho_c(E) = \begin{cases} \rho_0, & \frac{U}{2} - \frac{W}{2} \le |E| \le \frac{U}{2} + \frac{W}{2}, \\ 0, & \text{otherwise}, \end{cases} \tag{21}$$

where $U$ is the on-site Hubbard energy and $W$ is the chargon bandwidth. We write the chargon LDOS as $\rho_c(E, \mathbf{r}) = \rho_c(E) + \Delta\rho_c(E, \mathbf{r})$ where $\Delta\rho_c(E, \mathbf{r})$ represents the star-of-David spatial modulation superimposed on the constant background $\rho_c(E)$ and is modeled as

$$\Delta\rho_c(E, \mathbf{r}) = \rho_0 \sum_{i=1}^{6} \alpha_c(E, \mathbf{b}_i) \cos(\mathbf{b}_i \cdot \mathbf{r}), \tag{22}$$

where $\alpha_c(E, \mathbf{b}_i)$ is relative strength of the star-of-David CDW. Although $\alpha_c(E, \mathbf{b}_i)$ has strong energy dependence as pointed out earlier, here we take $\alpha_c(E, \mathbf{b}_i) \equiv \alpha_{\mathbf{c}}$ as a constant in the numerical simulation, which captures the essential physics and does not alter the main conclusions. Finally, the electron LDOS is obtained by convolution of these factors using Eq. (13).

The simulation results compared to the experimental data are summarized in Supplementary Fig. 16, where we have used a spinon density wave amplitude $\Delta_0 = 0.2t_\mathrm{s}$ which is only a small fraction of the spinon hopping $t_\mathrm{s}$ (~ spin-exchange *J*). The simulation results show that the electron density modulation at the spinon density modulation wavevectors $\mathbf{Q}_i$ persists over the entire Hubbard band energy range and has similar relative modulation strength (about 2%) to the experiment (2% ~6%). The theoretical ratio $r(E) \approx 1$ is also close to the experimental value $r(E) \approx 1 \sim 3$ (Fig. 4b).

**Supplementary Note 4: The short-range commensurate super-modulation (CS)**

A short-range commensurate super-modulation (CS) with $\sqrt{3} \times \sqrt{3}$ order is observed in the lower Hubbard band (LHB) in certain regions of single-layer 1T-$TaSe_2$/BLG (Fig. 3f). This pattern is reminiscent of short-range fluctuations arising from the 120° antiferromagnetic (AFM) order[20] in a triangular lattice.

The reason that STM can detect such spin density modulations can be understood using an argument similar to the discussion in Supplementary Notes 3.2 and 3.3. The coplanar 120° AFM order is described by spin order parameters $\mathbf{S}_{\mathbf{K}_i}$ ($\mathbf{K}_i$ labels all the K points at the corners of the first CDW BZ, where $1 \leqslant i \leqslant 6$) lying in the *x*-*y* plane. Such order parameters can be written in a complex form as in the X-Y model, where the real and imaginary parts correspond to the *x* and *y* polarization, respectively. The complex AFM order produces c-CDW order at wavevectors $\mathbf{K}_i \pm \mathbf{K}_j$ in way similar to how complex PDW order produces c-CDW order at higher harmonic wavevectors. Therefore, charge density super-modulations detected by STM appear at $\mathbf{K}_i$ (e.g. $\mathbf{K}_1 = \mathbf{K}_3 - \mathbf{K}_2$).

**Supplementary Figures**

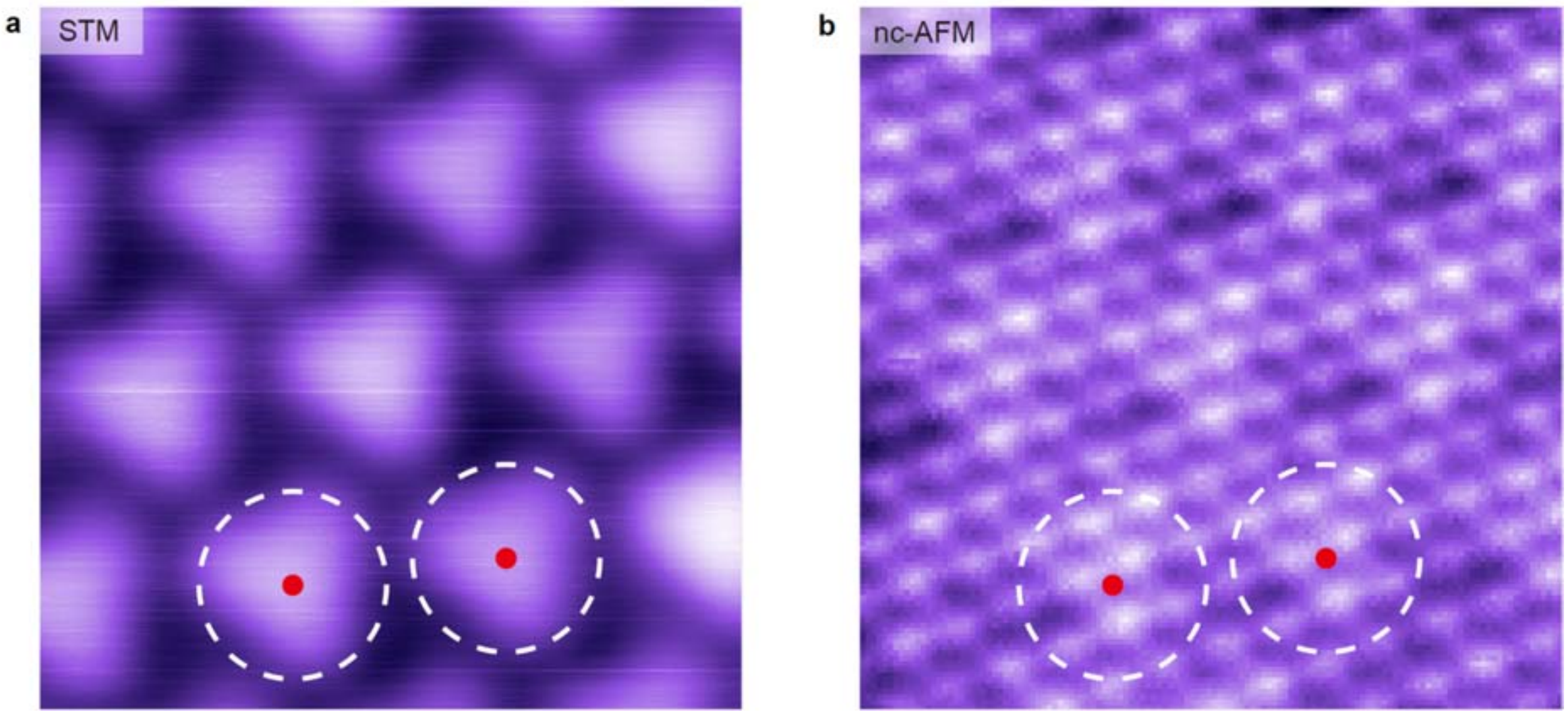

**Supplementary Fig. 1. The atomic structure of the star-of-David cell in single-layer 1T-$TaSe_2$/BLG revealed by STM and nc-AFM.**

**a**, Typical STM image of SL 1T-$TaSe_2$/BLG showing the star-of-David CDW patterns ($V_b$ = -0.5 eV, It = 10 pA). Two of the star-of-David cells are outlined by the white dashed circles and the star centers are marked by red dots. **b**, A nc-AFM frequency shift image acquired using constant-height mode in the same field of view as **a**. The top Se atoms are resolved. By comparing **a** and **b**, the center of each star-of-David cells (red dot) is seen to correspond to a Ta atomic site rather than a Se atomic site. The CDW structure of single-layer 1T-$TaSe_2$ thus shares a similar star-of-David structure to bulk 1T-$TaS_2$.

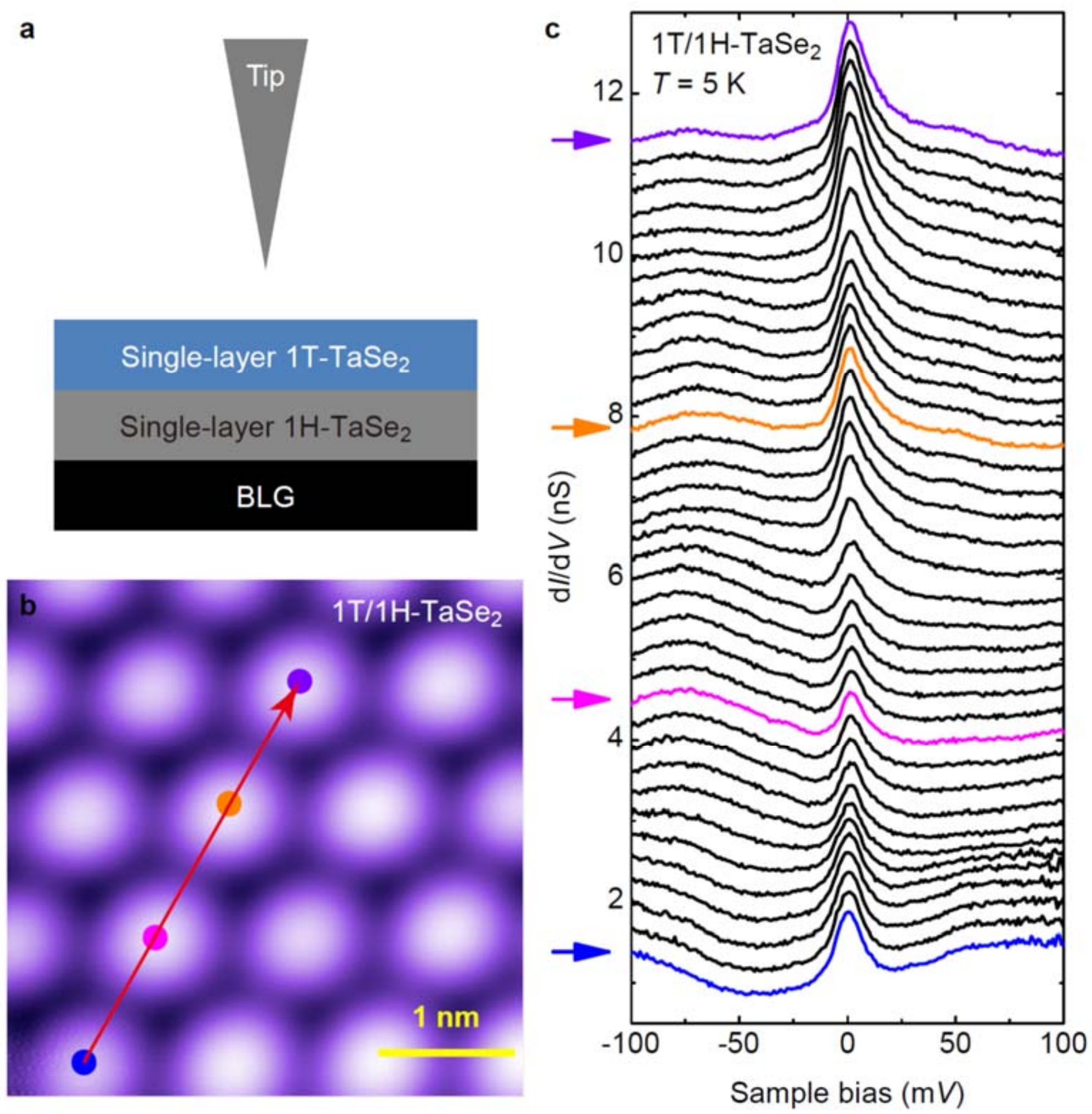

**Supplementary Fig. 2. Spatial dependence of the Kondo resonance in the 1T/1H-$TaSe_2$ vertical heterostructure on BLG/SiC(0001).**

**a**, Schematic of the 1T/1H vertical heterostructure. **b**, Typical topographic image of the top 1T layer in the 1T/1H vertical bilayer heterostructure shows a star-of-David CDW superlattice similar to that in single-layer 1T-$TaSe_2$ ($V_b$ = -0.1 V, $I_t$ = 30 pA). **c**, d*I*/d*V* spectra of the Kondo resonance peak acquired at different spatial locations along the red arrow in **b** ($V_b$ = -0.1 V, $I_t$ = 0.1 nA, $V_{mod}$ = 1 mV). The observation of the Kondo resonance peak at every star-of-David cell indicates the existence of a triangular lattice of spins in single-layer 1T-$TaSe_2$.

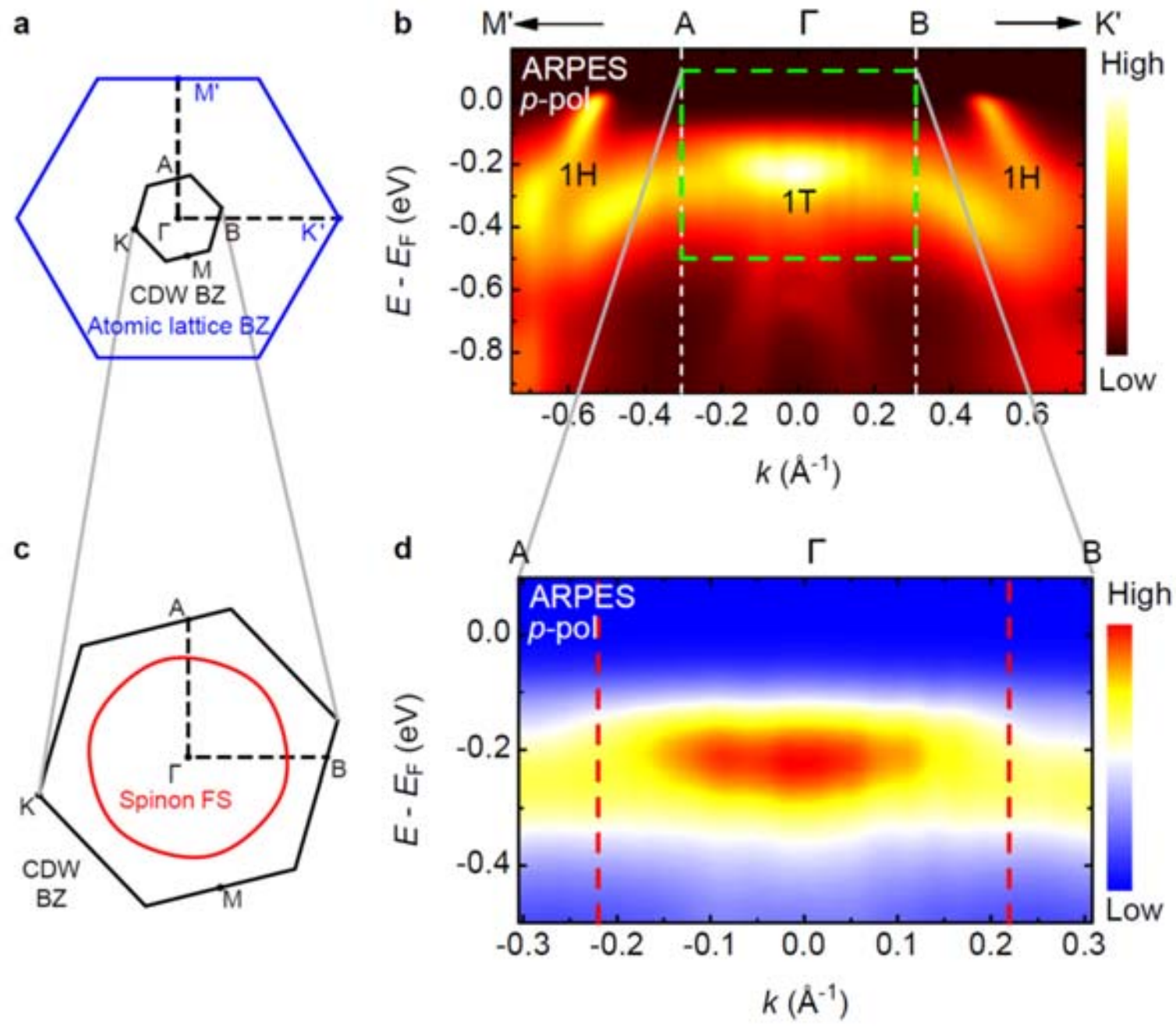

**Supplementary Fig. 3. ARPES of single-layer $TaSe_2$ on BLG/SiC(0001) at 12 K.**

**a**, The atomic lattice Brillouin zone (BZ) and the star-of-David CDW BZ of single-layer 1T-$TaSe_2$. **b**, ARPES data along M'-Γ-K' indicated by the black dashed lines in **a**. The two vertical white dashed lines indicate the star-of-David CDW BZ boundaries labeled as A and B. The green dashed rectangle around Γ highlights the lower Hubbard band (LHB) of single-layer 1T-$TaSe_2$, where the large broadening in energy (~ 0.1 eV) is consistent with a spectral continuum resulting from spin-charge separation (the broad spectral continuum could be due to other reasons such as the overlap of different bands in the DFT calculation[1]). The two bands crossing the Fermi level at large in-plane momenta are from single-layer 1H-$TaSe_2$, consistent with previous ARPES measurements[8]. Therefore, both the 1T and 1H single layers coexist. **c**, The star-of-David CDW BZ and the calculated spinon FS (see Materials and Methods). **d**, Zoomed-in ARPES data within the green dashed rectangle in **b** shows the LHB of single-layer 1T-$TaSe_2$. The two vertical red dashed lines indicate the calculated spinon Fermi wavevectors ($k_F$) in both directions. The broadened LHB spectral weight decays rapidly beyond $k_F$, consistent with the existence of a spinon Fermi surface.

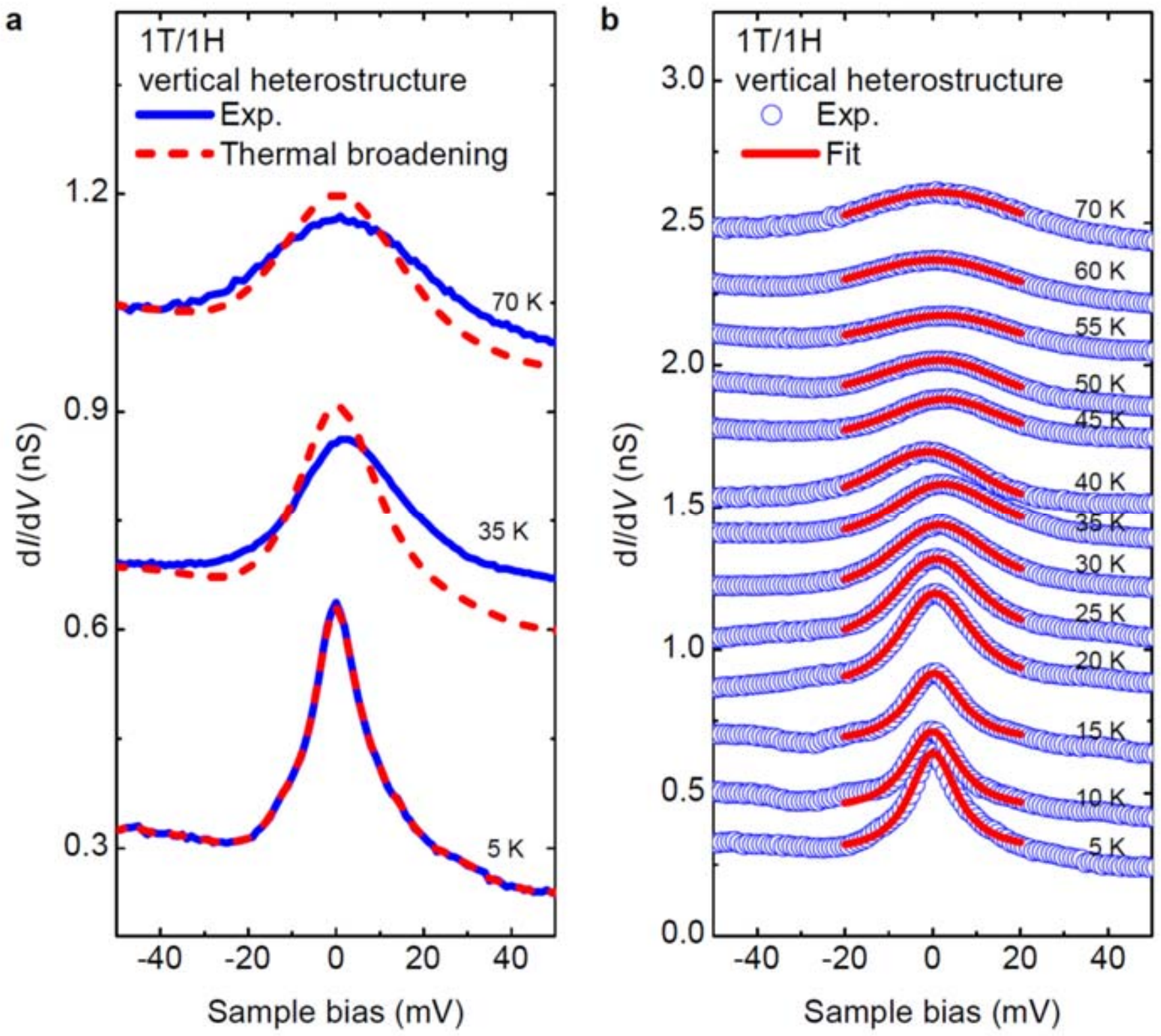

**Supplementary Fig. 4. Thermal broadening simulation and thermally convolved Fano fit of the temperature-dependent Kondo resonance peak in the 1T/1H-$TaSe_2$ vertical heterostructure.**

**a**, Thermal broadening simulation (red dashed lines) of the temperature-dependent Kondo resonance d*I*/d*V* spectra at 5 K, 30 K, and 70 K (blues lines, replotted from Fig. 2b). Here the zero-temperature density of states is approximated by the experimental d*I*/d*V* spectrum at 5 K, which is then thermally convolved with thermal broadening factors at elevated temperatures to give the simulated curves (see Materials and Methods for details). The experimental Kondo resonance peak, being broader than the simulation, cannot be accounted for by purely thermal broadening. **b**, Thermally convolved Fano fits (red lines) of the Kondo resonance d*I*/d*V* spectra at all measured temperatures. The fits are performed by thermally convolving the Fano line shape plus a constant background with a thermal broadening factor (see Materials and Methods for details).

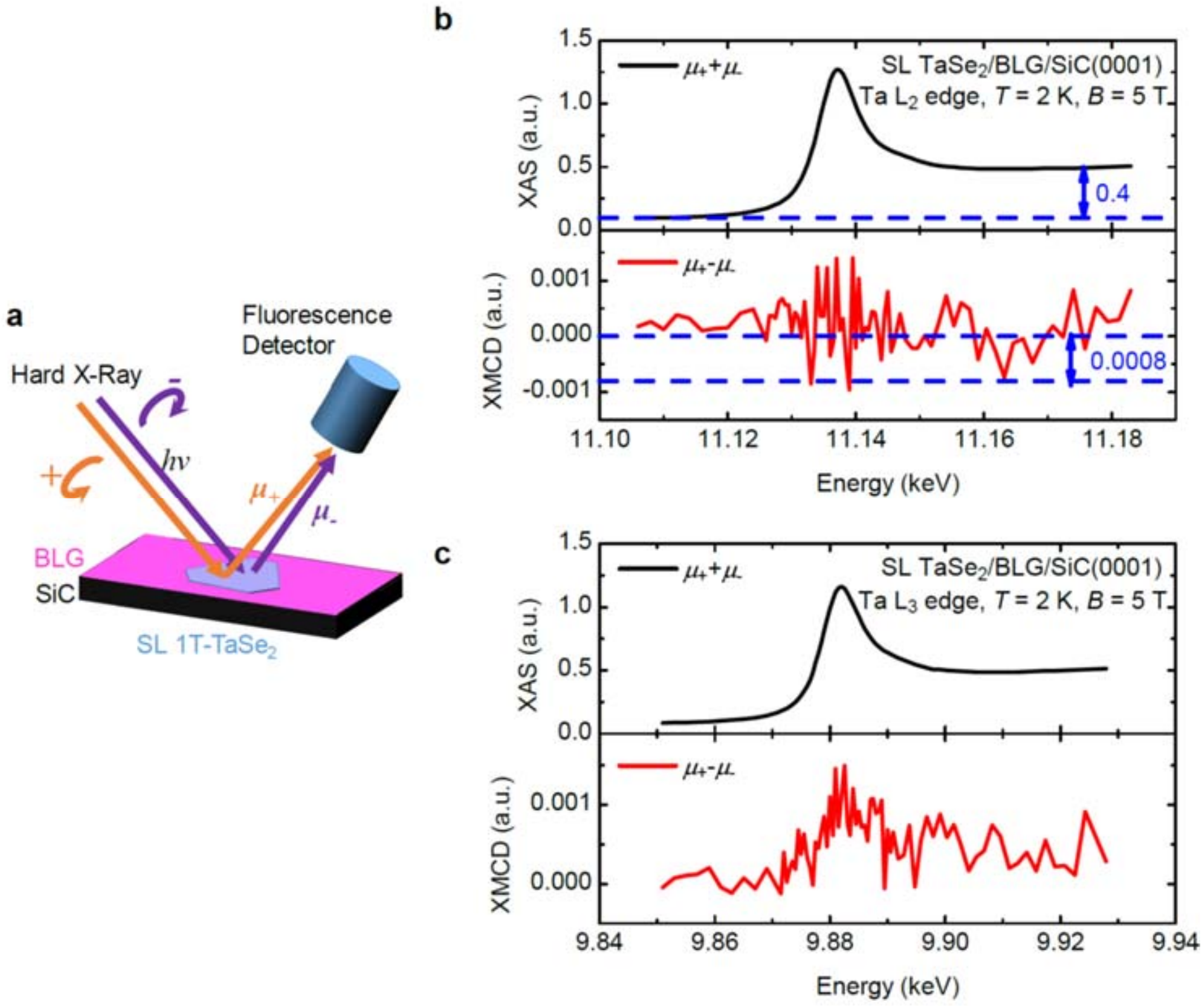

**Supplementary Fig. 5. X-ray magnetic circular dichroism (XMCD) spectra of single-layer 1T-$TaSe_2$/BLG measured at Ta $L_2$/$L_3$ edges ($B$ = 5 T, $T$ = 2 K).**

**a**, Schematic of XMCD measurement of SL 1T-$TaSe_2$/BLG. The X-ray absorption spectra (XAS) were collected by measuring the fluorescence signal for both right circularly polarized light ($\mu_+$) and left circularly polarized light ($\mu_-$). **b**, Total X-ray absorption ($\mu_+ + \mu_-$, black curve) and XMCD ($\mu_+ - \mu_-$, red curve) of Ta $L_2$ edge. There is no observable XMCD signal. The upper limit of XMCD is set by the noise level ~ 0.0008 a.u., which is ~ 0.2% if normalized by the step height ~0.4 a.u. across the $L_2$ edge. This gives an estimated upper limit for the magnetic moment of ~ 0.013 $\mu_B$ per star-of-David cell by using the results of Ref.[21] where an induced Ta 5*d* magnetic moment in a CoCrPtTa alloy was estimated to be 0.035 $\mu_B$ per Ta atom from a 4% Ta $L_2$ XMCD signal. **c**, Total X-ray absorption ($\mu_+ + \mu_-$, black curve) and XMCD ($\mu_+ - \mu_-$, red curve) of Ta $L_3$ edge. There is no observable XMCD signal either.

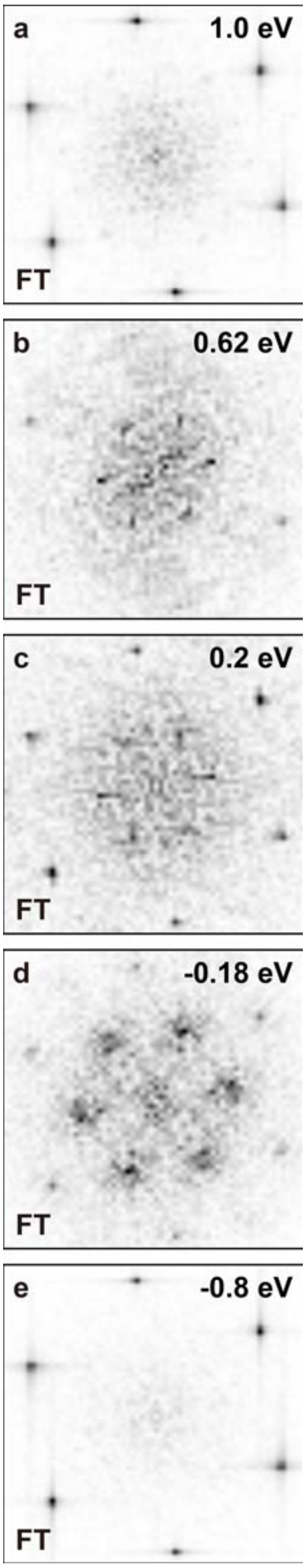

**Supplementary Fig. 6. Fourier transform images replotted from Fig. 3h-l without labels for clearer inspection.**

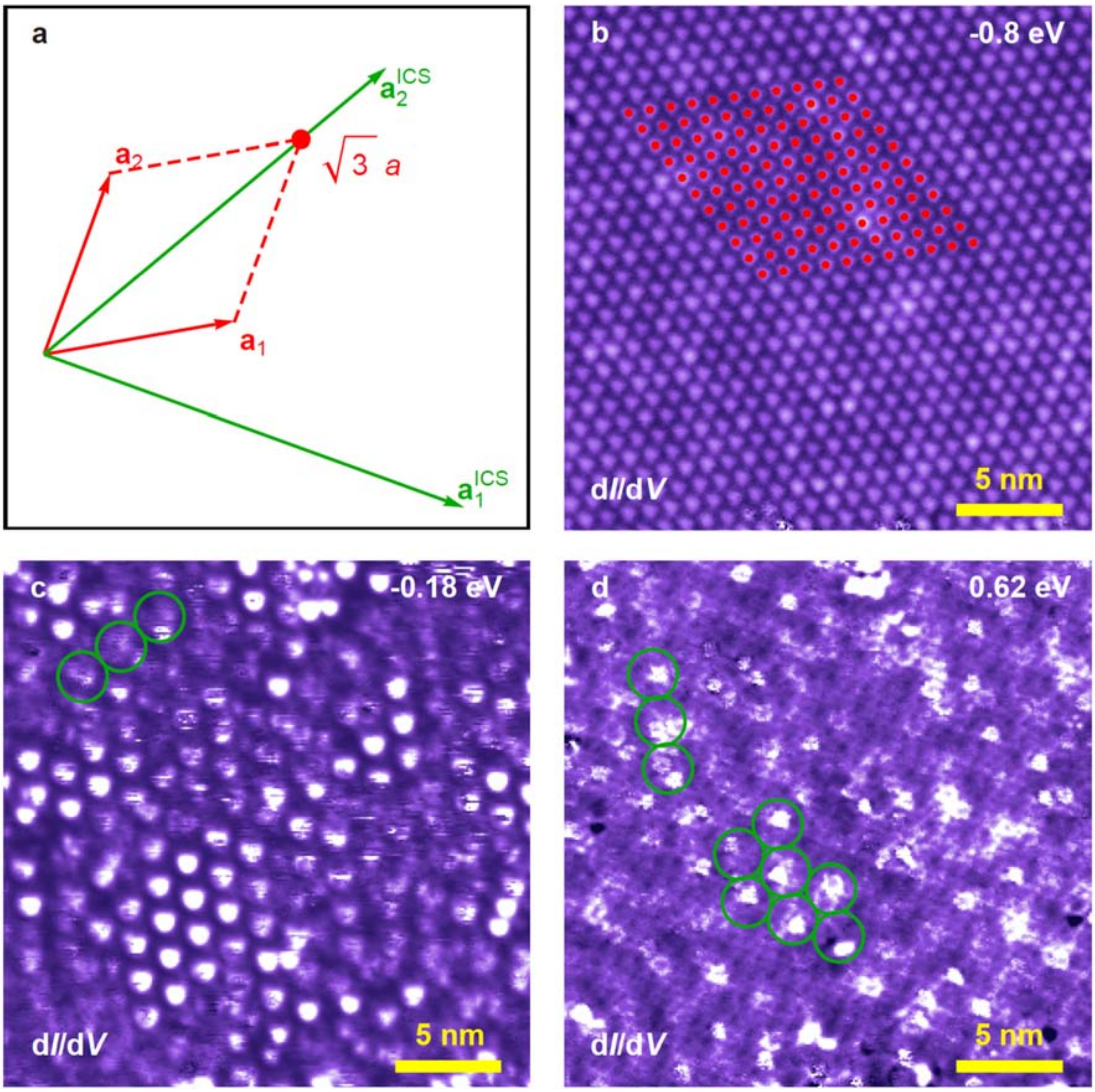

**Supplementary Fig. 7. Real-space registration of the ICS pattern in SL 1T-TaSe$_2$/BLG.** **a**, Schematic of real-space unit vectors of the star-of-David CDW lattice (red) and the ICS pattern (green). Analysis of **b-d** indicates that the ICS pattern is rotated from the CDW lattice by 30° and has a wavelength larger than the commensurate value $\sqrt{3}a$, where $a$ is the CDW lattice constant. **b**, Registration of the CDW lattice (red dots) in a selected area of the d$I$/d$V$ map in the filled states ($V_b$ = -0.8 V). **c**, Registration of the ICS pattern (green circles) in a selected area of the d$I$/d$V$ map in the LHB ($V_b$ = -0.18 V). **d**, Registration of the ICS pattern (green circles) in selected areas of the d$I$/d$V$ map in the UHB$_2$ ($V_b$ = 0.62 V).

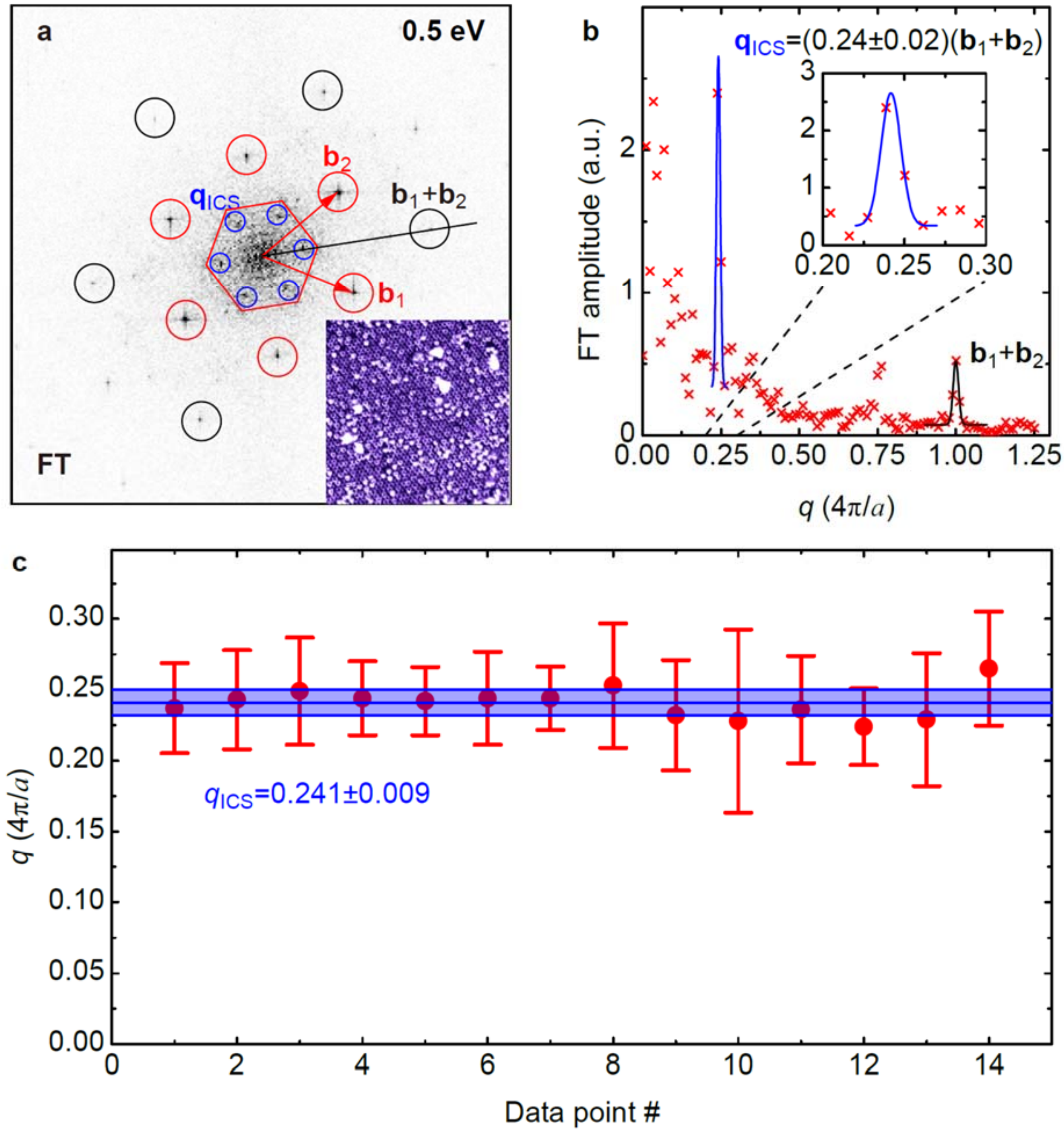

**Supplementary Fig. 8. Measuring the ICS wavevector in single-layer 1T-$TaSe_2$/BLG.** **a**, Fourier transform (FT) of the 50 nm × 50 nm topographic image shown in bottom right inset ($V_b$ = 0.5 V, $I_t$ = 2 pA). The red hexagon represents the CDW Brillouin zone (BZ). The CDW wavevectors (red circles) and their higher harmonics (black circles), as well as the ICS wavevectors (blue circles), are all resolved in the FT image. The topographic image was acquired at $V_b$ = 0.5 V in the $UHB_2$ to enhance the ICS feature by integrating contributions from the whole $UHB_1$. **b**, FT amplitude as a function of wavevector along the Γ-K direction

indicated by the black line in **a**. The $\mathbf{b}_1 + \mathbf{b}_2$ and ICS peaks are fitted by Gaussian line shapes (black and blue lines, respectively). The inset displays the ICS peak and its Gaussian fit over a zoomed-in *q*-range of $0.2 \leqslant q \leqslant 0.3$. The fit yields an ICS wavevector $\mathbf{q}_{\mathrm{ICS}} = (0.24 \pm 0.02)(\mathbf{b}_1 + \mathbf{b}_2)$. **c**, The ICS wavevector $\mathbf{q}_{\mathrm{ICS}}$ measured on different samples using different STM tips (red dots with error bars) and performing similar Gaussian fits as explained in **a** and **b**. This procedure yields an average value of $\mathbf{q}_{\mathrm{ICS}} = (0.241 \pm 0.009)(\mathbf{b}_1 + \mathbf{b}_2)$ (blue).

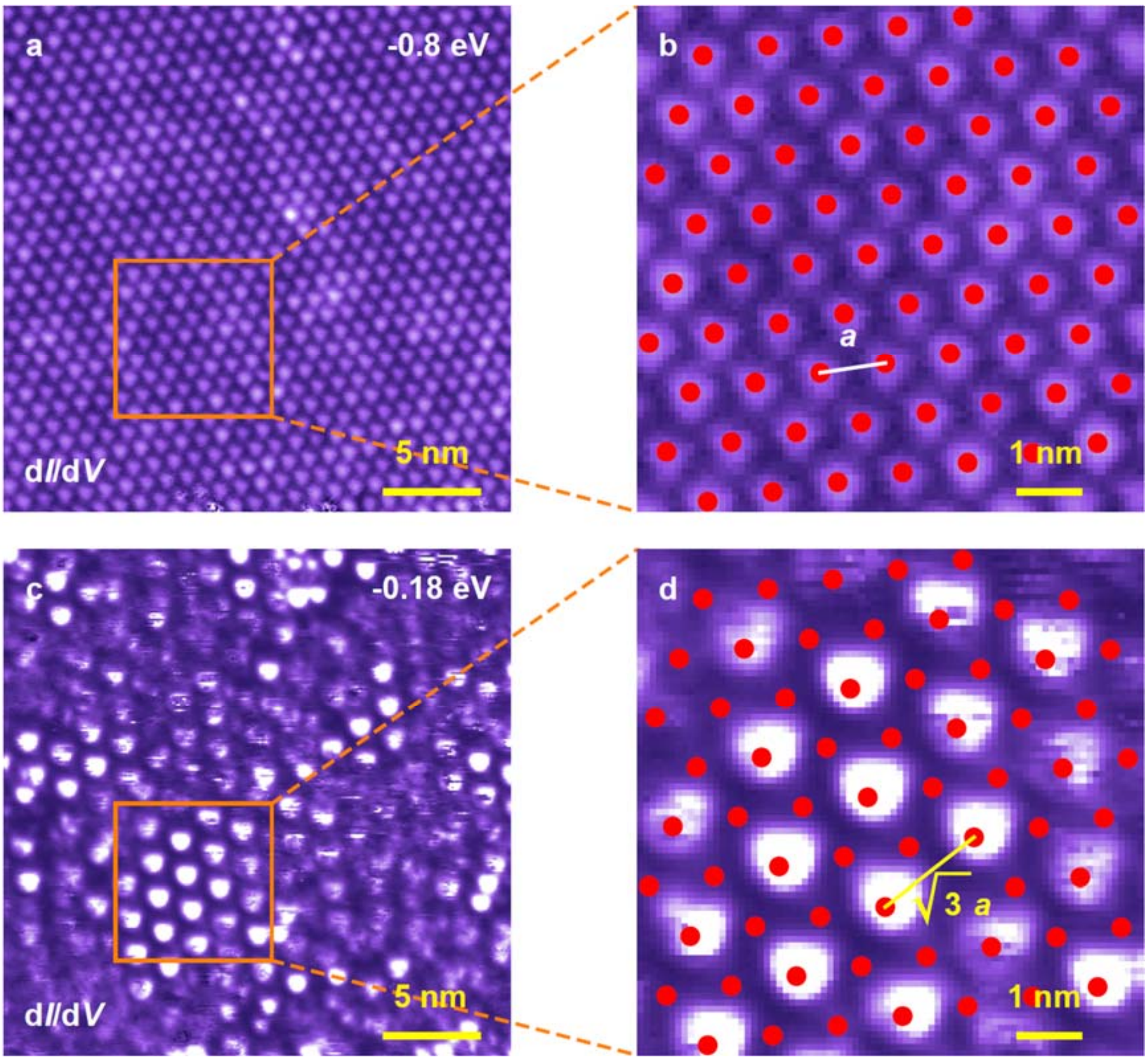

**Supplementary Fig. 9. Real-space registration of the CS pattern in SL 1T-$TaSe_2$/BLG.** **a**, d*I*/d*V* map in the filled state showing the star-of-David CDW pattern ($V_b$ = -0.8 V, replotted from Fig. 3g). **b**, Zoomed-in d*I*/d*V* map within the orange square in **a**. The CDW lattice is indicated by an array of red dots and the CDW lattice constant is $a$. **c**, d*I*/d*V* map in the LHB showing both the ICS and CS patterns in the same area as **a** ($V_b$ = -0.18 V, replotted from Fig. 3f). The short-range CS pattern appears as patches of bright spots, and one such patch is enclosed in the orange square. **d**, Zoom-in of the d*I*/d*V* map in **c** within the orange square (the same area as in **b**). The CDW lattice is indicated by the red dots as determined in **b**. Compared to the CDW lattice, the short-range CS pattern is rotated by 30° and has a commensurate wavelength of $\sqrt{3}a$.

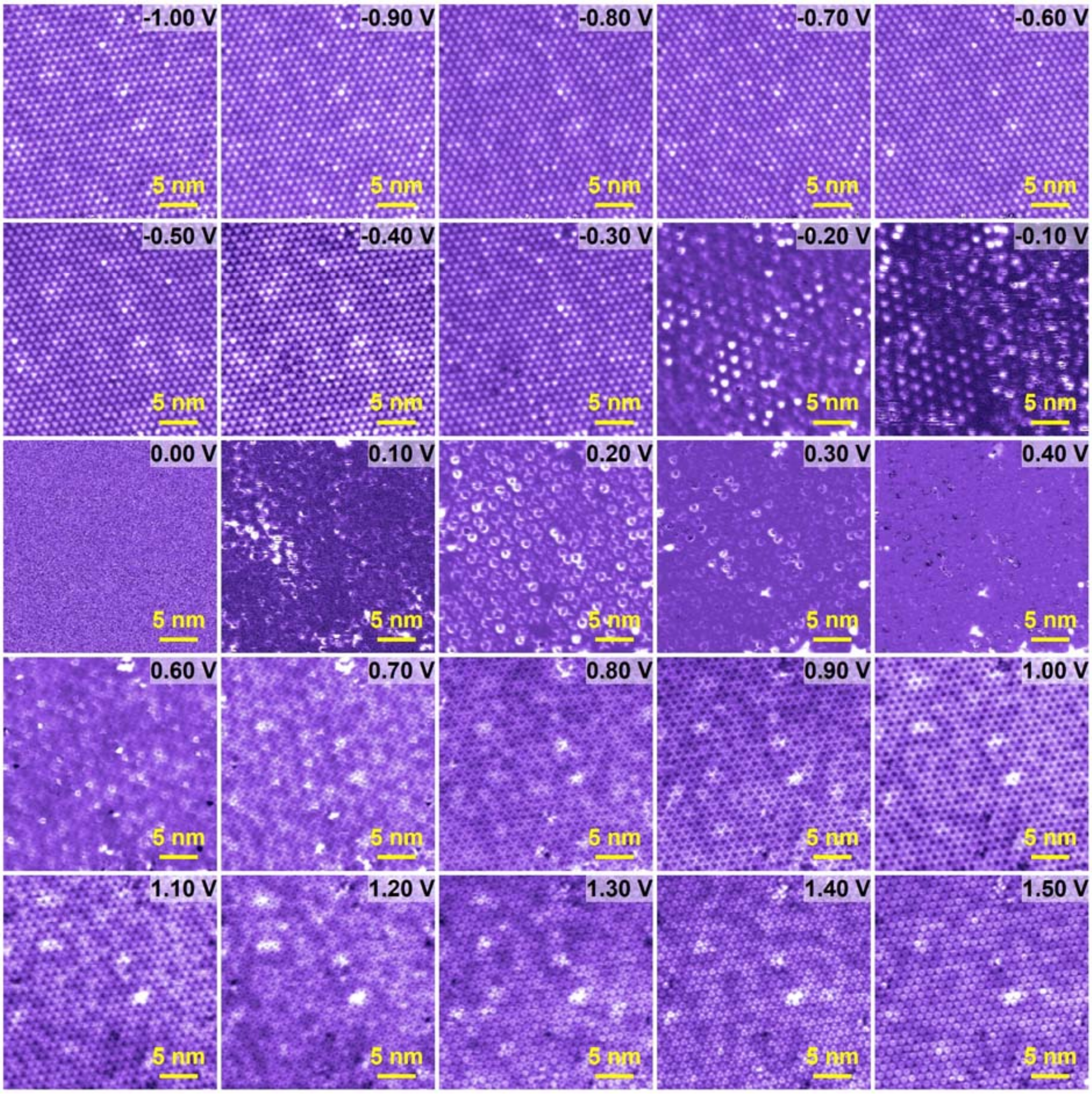

**Supplementary Fig. 10. Differential conductance maps in single-layer 1T-$TaSe_2$/BLG at energies -1.0 eV ⩽ $E$ ⩽ 1.5 eV.**

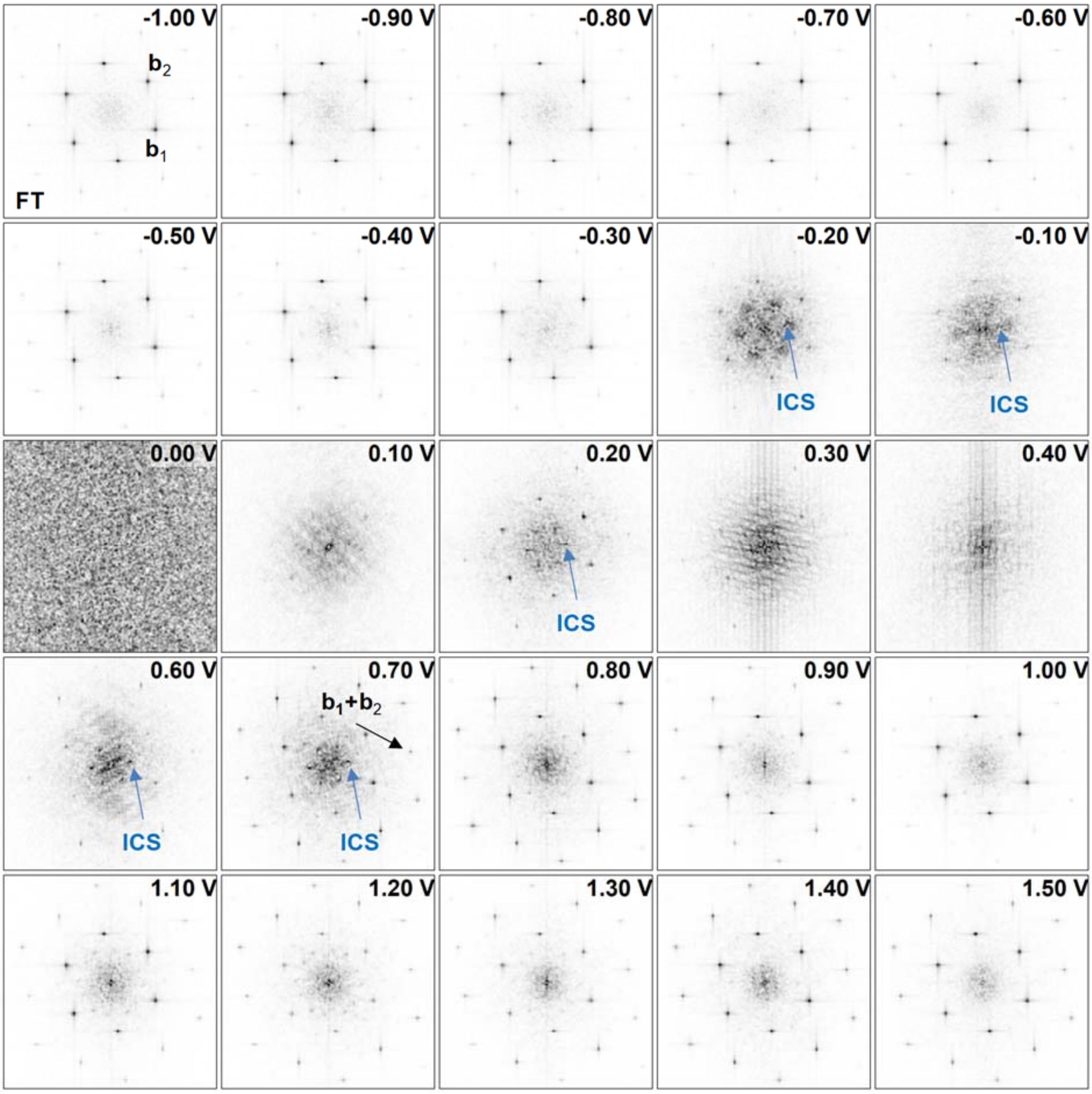

**Supplementary Fig. 11. Fourier transform images of the conductance maps shown in Supplementary Fig. 10.**

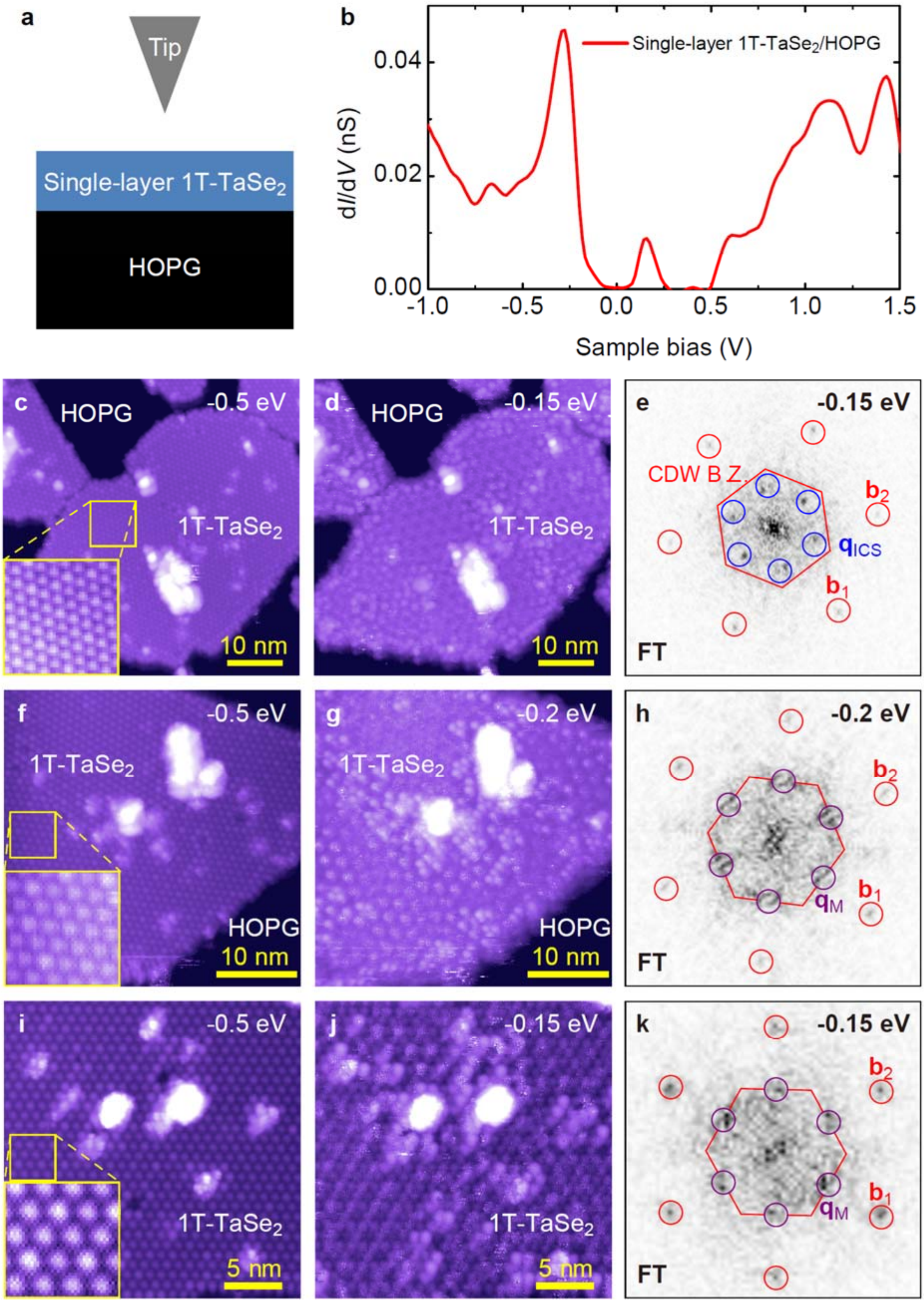

**Supplementary Fig. 12. Super-modulations in single-layer 1T-$TaSe_2$/HOPG.**

**a**, Schematic of single-layer 1T-$TaSe_2$ grown on HOPG substrate via MBE. **b**, Typical d$I$/d$V$

spectrum of single-layer 1T-$TaSe_2$/HOPG shows the same Mott insulating behavior as in single-layer 1T-$TaSe_2$/BLG. **c**, Topographic image shows the star-of-David CDW ($V_b$ = -0.5 V, $I_t$ = 1 pA). **d**, Topographic image shows the ICS pattern in the same area as in **c** ($V_b$ = -0.15 V, $I_t$ = 1 pA). **e**, Fourier transform (FT) of **d**. The CDW reciprocal lattice unit vectors, the CDW Brillouin zone (BZ), and the ICS wavevectors are marked by red circles, red hexagon, and blue circles, respectively. **f**, Topographic image shows the star-of-David CDW on a different island ($V_b$ = -0.5 V, $I_t$ = 1 pA). **g**, Topographic image shows a 2 × 2 super-modulation with respect to the CDW lattice in the same area as shown in **f** ($V_b$ = -0.2 V, $I_t$ = 1 pA). **h**, FT of **g** shows FT peaks of the 2 × 2 super-modulation located at $\mathbf{q}_M$ wavevectors (purple circles) close to the M points of the CDW BZ. **i-k**, Another dataset on a different island shows the star-of-David CDW and the 2 × 2 $\mathbf{q}_M$ super-modulation ($I_t$ = 1 pA).

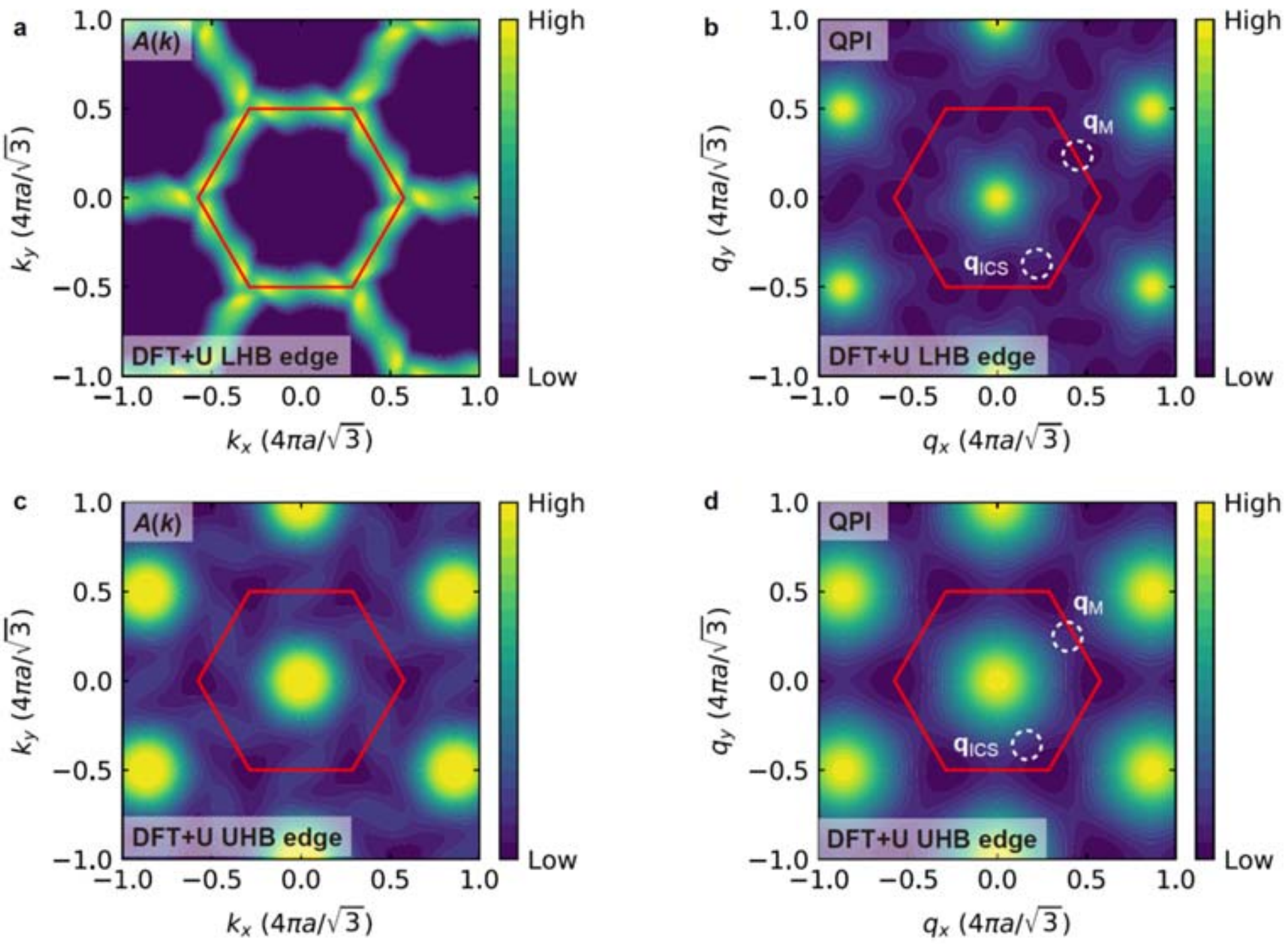

**Supplementary Fig. 13. Simulation of electronic quasiparticle interference (QPI) pattern in SL 1T-TaSe$_2$ using DFT+*U*.**

**a**, The momentum-space map of the electron spectral function at the LHB edge calculated by DFT+*U* (details of the DFT+*U* calculation are described in ref.[1]). The red hexagon is the first Brillouin zone (BZ) of the star-of-David CDW lattice. **b**, Simulated electronic QPI pattern at the LHB edge obtained by calculating the joint density of states (DOS)[22] from the DFT+*U* band structure shown in **a**. The dashed white circles represent the *q* vectors where the $\mathbf{q}_{\mathrm{ICS}}$ and $\mathbf{q}_{\mathrm{M}}$ super-modulations should be seen. There are no clear features at these wavevectors in the simulation. **c**, The momentum-space map of the electron spectral function at the UHB edge calculated by DFT+*U*. **d**, Simulated electronic QPI pattern at the UHB edge obtained by calculating the joint DOS from the DFT+*U* band structure shown in **c**. There are no clear features at the $\mathbf{q}_{\mathrm{ICS}}$ and $\mathbf{q}_{\mathrm{M}}$ wavevectors in the simulation. This indicates that the experimentally observed super-modulations at $\mathbf{q}_{\mathrm{ICS}}$ and $\mathbf{q}_{\mathrm{M}}$ cannot be explained by electronic QPI at this level of theory.

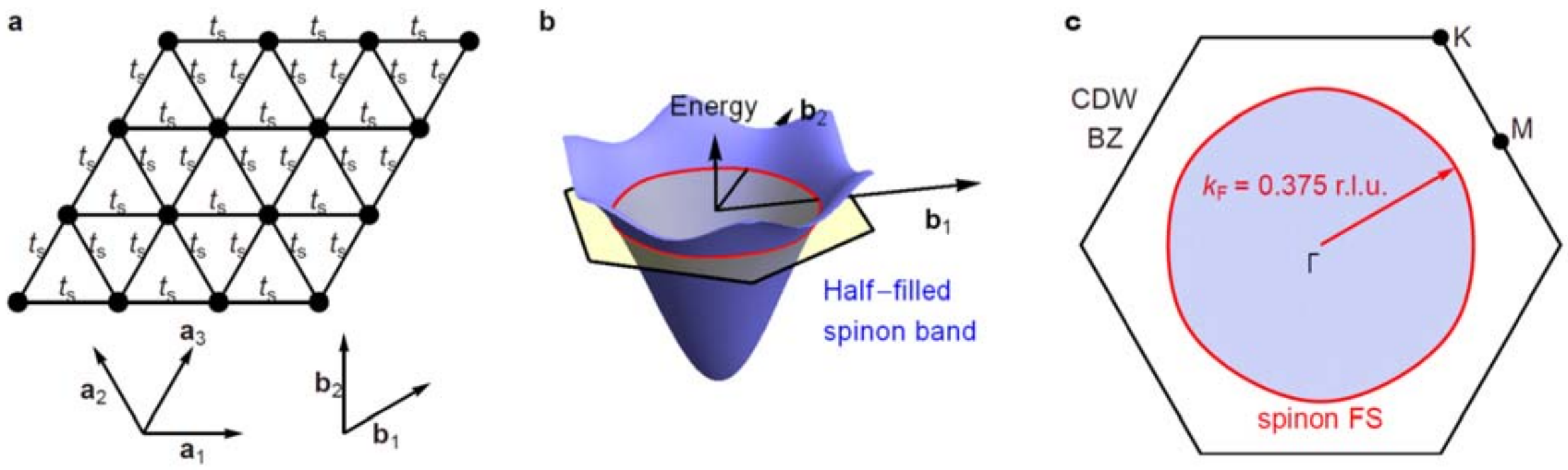

**Supplementary Fig. 14. Simulation of the spinon band structure using a tight-binding model.**

**a**, Schematic of the zero-flux mean-field tight-binding model for the spinons in a triangular lattice (see Materials and Methods). All nearest-neighbor hopping amplitudes are taken to be the same ($t_s$). **b**, Calculated spinon band structure from the tight-binding model. The CDW Brillouin zone (BZ) is represented by the black hexagon. The Fermi surface (FS) of the half-filled spinon band is marked red. **c**, The spinon FS (red) is shown in two-dimensional momentum space. The CDW BZ is represented by the black hexagon. The spinon FS of the half-filled spinon band is represented by the slightly distorted red circle that encloses exactly half the area of the whole CDW BZ. The Fermi wavevector along the Γ-M direction is $k_F \approx$ 0.375 r.l.u. (r.l.u. = reciprocal lattice unit).

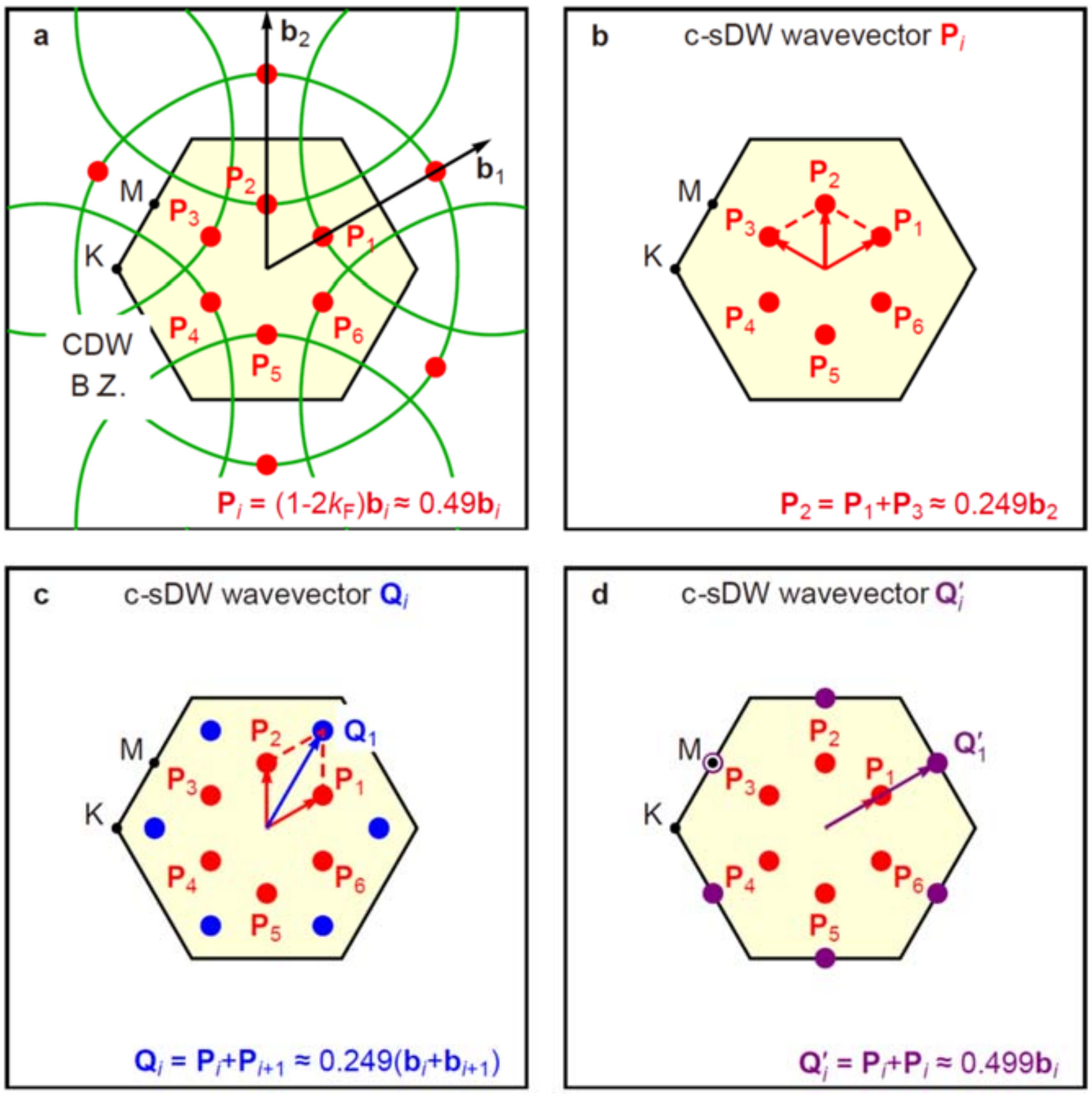

**Supplementary Fig. 15. Predicted spinon FS instability wavevectors and composite spinon density wave (c-sDW) wavevectors at higher harmonics.**

**a**, The predicted spinon FS instability wavevectors are shown as red dots. The black hexagon represents the first Brillouin zone (BZ) of the CDW superlattice, and the $\mathbf{b}_i$'s are the CDW reciprocal lattice unit vectors. The green lines are the $2k_F$ surface lines obtained from the tight-binding simulation (Supplementary Fig. 14). Previous DMRG simulations[23] predict spinon FS instabilities occurring in the Γ-M directions at the red dots on the $2k_F$ surface lines. The red dots within the first CDW BZ are labeled by $\mathbf{P}_i$'s ($1 \leqslant i \leqslant 6$). The composite sDW (c-sDW) wavevectors can be obtained as the higher harmonics of the primary wavevectors $\mathbf{P}_i$'s (see Supplementary Notes 3.2 and 3.3). **b**, The c-sDW wavevectors $\mathbf{P}_i = \mathbf{P}_{i-1} + \mathbf{P}_{i+1}$ (e.g. $\mathbf{P}_2 = \mathbf{P}_1 + \mathbf{P}_3$). **c**, The c-sDW wavevectors $\mathbf{Q}_i = \mathbf{P}_i + \mathbf{P}_{i+1}$ marked by blue dots. **d**, The c-sDW wavevectors $\mathbf{Q}_i' = 2\mathbf{P}_i$ marked by purple dots are very close to the M points.

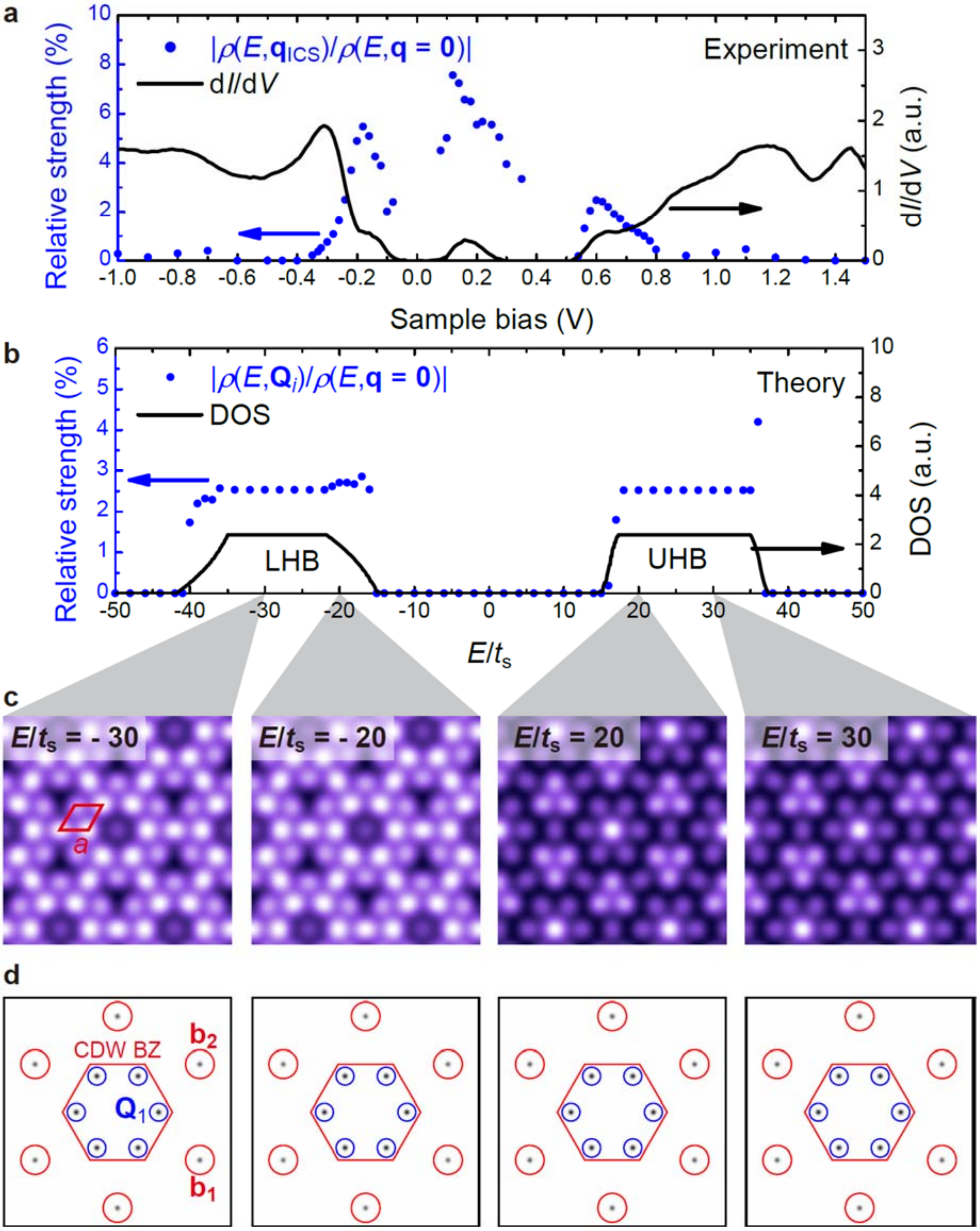

**Supplementary Fig. 16. Simulation of spinon and chargon spectral function convolution compared with experiment.**

**a**, Experimental energy dependence of the ratio of the ICS strength $\rho(E, \mathbf{q}_{\text{ICS}})$ with respect to the spatially averaged LDOS $\rho(E, \mathbf{0})$ (the non-oscillating constant background). The relative

ICS strength is about 2% ~ 6% of the average background. **b**, Simulated energy dependence of the relative ICS strength with respect to the non-oscillating constant background. The relative strength is about 2% and the ICS extends over the entire Hubbard bands in the simulation, consistent with the experiment. The simulation method is detailed in Supplementary Note 3.4. Here the spinon Brillouin zone was sampled by a $128 \times 128$ mesh and the other model parameters were as follows: the spinon energy resolution was $\Delta E_\mathrm{s} = 0.1 t_\mathrm{s}$, the spinon density wave amplitude was $\Delta_0 = 0.2 t_\mathrm{s}$, the spinon energy Gaussian smearing factor was $\sigma = 0.1 t_\mathrm{s}$, the chargon bandwidth was $W = 20 t_\mathrm{s}$ (larger than the spinon bandwidth $9 t_\mathrm{s}$), the Hubbard $U$ was $U = 50 t_\mathrm{s}$, and the relative strength of the chargon star-of-David pattern to the spatially averaged chargon DOS was $\alpha_\mathrm{c} = 0.02$. **c**, Simulated electron LDOS in the LHB and UHB showing both the ICS pattern and the star-of-David CDW. The star-of-David CDW unit cell is marked by a red diamond where $a$ is the CDW lattice constant. **d**, Fourier transform of the simulated electron LDOS in **c**. The red hexagon represents the first Brillouin zone (BZ) of the CDW superlattice, and $\mathbf{b}_1$, $\mathbf{b}_2$ are the CDW reciprocal lattice unit vectors (circled in red). The ICS wavevectors are circled in blue.